\documentclass[preprintnumbers,reprint,twocolumn, amsmath,amssymb, aps]
{revtex4-1}

\usepackage{graphicx}
\usepackage{dcolumn}
\usepackage{bm}
\usepackage{float}
\usepackage{hyperref}
\usepackage{subfig}
\usepackage{dsfont}
\usepackage{slashed}
\usepackage{color}
\usepackage{amsmath}
\usepackage[section]{placeins}
\usepackage{braket}
\usepackage{upgreek}
\usepackage[normalem]{ulem}

\newcommand{\qqb}{Q\bar{Q}}

\newcommand{\dbar}{d\hspace*{-0.08em}\bar{}\hspace*{0.1em}}

\begin{document}
\preprint{TIFR/TH/22-xx}

\title{Energy hierarchies governing quarkonium dynamics in Heavy Ion Collisions}

\author{Rishi Sharma}
\email{rishi.sharma@gmail.com}
 \affiliation{Tata Institute of Fundamental Research \\
 Colaba, Mumbai, India. 400005
 }
\author{Balbeer Singh}
 \email{balbeer@theory.tifr.res.in}
\affiliation{
 Tata Institute of Fundamental Research \\
 Colaba, Mumbai, India. 400005
}

\begin{abstract}
In this paper, we critically examine hierarchies between energy scales that
determine quarkonium dynamics in the quark gluon plasma. A particularly
important role is played by the ratio of the binding energy of species ($E_b$)
and the medium scales; temperature ($T$) and Debye mass ($m_D$). It is well
known that if these ratios are much larger than one then the dominant process
governing quarkonium evolution is dissociation by thermal gluons
(gluo-dissociation). On the other hand, if this ratio is much smaller than one
then quarkonium dynamics is dominated by screening and Landau damping of the
exchanged gluons. Here we show that over most of the evolution, the scale
hierarchies do not fall in either limit and one needs to use the full structure
of the gluonic spectral function to follow the dynamics of the $Q\bar{Q}$ pair.
This has a significant bearing when we follow the quantum dynamics of quarkonia
in the medium. The inverse medium relaxation time  is also $\sim T$ and if
$E_b$ is comparable (or larger) in magnitude to $T$,  the quantum evolution of
$Q\bar{Q}$ is non-local in time within the Brownian approximation. 
\end{abstract}

\maketitle

\section{\label{sec:introduction}Introduction}

In the vacuum, the  bound states  of heavy quarks ($Q$ which can be $b$ or $c$)
and anti-quarks ($\bar{Q}$) feature three prominent momentum scales:
the heavy quark masses $M$, the inverse relative separation $\frac{1}{r}$,
and the binding energies $E_b$ (see Ref.~\cite{brambilla20001} for a
comprehensive review). These scales
satisfy the hierarchies $M\gg \frac{1}{r} \gg E_b$, which justifies
non-relativistic treatments of these states. An additional relevant scale
for their description is the scale of quantum Chromodynamics ($\Lambda_{QCD}$)
which may or may not be somewhat smaller than $E_b$~\cite{brambilla20001} but can
be assumed to be significantly smaller than $1/r$ (especially for $\bar{b}b$ pair).
The hierarchy of these scales allows one to integrate out modes at the scale $M$,
and $1/r$ systematically, and derive a low energy effective field theory (EFT) 
valid at the scale $E_b$. At the
lowest order in $rE_b$, the EFT consists of non-relativistic quarks bound by a
potential~\cite{pineda19981}. At higher order, the theory
features interactions mediated by
gluons of wavelength $1/E_b$. Effects of higher order terms are suppressed by positive
powers of $rE_b$, where factors of $r$ can be seen as arising from a long wavelength
expansion of the fields. This framework is called pNRQCD~\cite{brambilla20001}.

In a thermal medium at temperature $T$, new scales appear which govern the dynamic
properties of quarkonia in the medium. It was pointed out in a classic paper
~\cite{matsui19861} that the screening of the $Q\bar{Q}$ interaction on an
inverse length scale $m_D$ could lead to the ``melting'' of the
bound states. Moreover, later on, it was realized that scattering between bound
states and the thermal constituents of the medium plays a major role in the
dissociation of the quarkonium states. This leads to the generation of the
imaginary part of the quarkonium potential in a thermal
medium~\cite{laine20071}.  Additionally, absorption of thermal gluons in the
medium could lead to gluo-dissociation~\cite{bhanot19791}.  It was
shown~\cite{brambilla20081} that pNRQCD naturally incorporates processes
leading to gluo-dissociation~\cite{peskin19791,bhanot19791} as its dynamical
degree of freedom that includes low energy gluonic degrees of freedom (and other
light degrees of freedom if any) in addition to the wavefunctions of $\qqb$
pair.  The corresponding emergent scale $\Gamma\sim {1}/{\tau_R}$ (where
$\tau_R$ is the relaxation time of quarkonia) is related to the dynamics of
inelastic interactions of $\bar{Q}Q$ with the medium.  Furthermore, $E_b$ and
$\frac{1}{r}$ might themselves be modified from their vacuum values by these
thermal effects.

Two medium scales that play a role in quarkonium dynamics in the quark gluon plasma (QGP) are
$T$ and $m_D$. In the weak coupling limit, there is a hierarchy between $m_D$ and
$T$~\cite{Braaten:1990it}. In this regime the coupling $g$ is small and
$m_D\sim gT\ll T$. However, for the temperatures of interest ($150-500$ MeV),
lattice results suggest that $m_D/T\sim 2$~\cite{kaczmarek20031}. Using $2\pi T$ as
the relevant energy scale~\cite{Brambilla:2020qwo} at which 
$\alpha_s$ is computed also gives similar values of $g$. 

This implies that in this regime leading order weak coupling expressions
in $g$ are not quantitatively reliable. For example, it is known that
the higher order corrections to the momentum diffusion coefficient are larger
than the leading order value~\cite{moore20081}.  However, non-perturbative calculations of some relevant 
dynamical processes is still challenging and weak-coupling calculations are still useful. An important
result in weak-coupling was obtained in Ref.~\cite{laine20071} which showed
that the potential between quark-antiquark pair is complex at finite $T$.

Such weak-coupling calculations have given insight into the problem and results
from these calculations can be used to obtain estimates for experimental
observables of interest: for example $R_{AA}$ in heavy ion collisions (HIC).

Many such calculations have been attempted to address the phenomenology of
quarkonium states in the QGP (see Ref.~\cite{andronic20151} for a review). For
approaches using a medium-modified $T$-matrix approach see
Refs.~\cite{grandchamp20041,rapp20091,zhao20111,emerick20111,zhao20121,du20171,du20181}.
Gluo-dissociation as the dominant mechanism for dissociation has been used in
Refs.~\cite{brezinski20121,nendzig20131,hong20191}. For approaches based on the
complex potentials derived by~\cite{laine20071} see
Refs.~\cite{strickland20111,strickland20112,margotta20111,krouppa20151,krouppa20171,krouppa20181}.
For approaches based on Schr\"{o}dinger-Langevin equation see
Refs.~\cite{Katz:2015qja,Gossiaux:2016htk,Bernard:2016spw}. Quarkonia at high
$p_T$ have been explored in Refs.~\cite{sharma20121,aronson20171,makris20191}. For
quantum dynamics in weak coupling see Refs.~\cite{akamatsu20181,brambilla20171,brambilla20181,Islam:2020bnp,Sharma:2019xum}. 

In this paper, we will focus on bottomonia and use leading order expressions
for the gluon polarization tensor. But we will not assume $m_D/T$ is small.
While this is not a formal expansion in $g$ but might better capture some
important qualitative dynamical properties of the QGP. This has been used in
other papers for open heavy flavor~\cite{moore20051}. 

The next question is how the thermal scales compare to the scales associated
with bound states. Clearly, $M\gg T$ and a non-relativistic treatment is
applicable for quarkonia slowly moving in the medium. For bottomonia, the
values of $1/r$ are comparable to $\sim1$GeV~\cite{ymocs20071} and we will
assume that $1/r\gg T$ and hence use pNRQCD to describe the system. 
However, we do not assume that the hierarchy between $1/r$ and the screening mass $m_D$ is so strong that we can ignore the screening of the $Q\bar{Q}$ 
potential when calculating quarkonium properties in the medium. In practice, we see that the effect of screening on the $\Upsilon(1{\rm{S}})$ wavefunction is small, but the $\Upsilon(2{\rm{S}})$ and $\Upsilon(3{\rm{S}})$ states are affected by screening. 

On the other hand, there is no clear separation between $E_b$ and $m_D,
T$ (FIG.~\ref{be} below). Moreover, their relative order depends on the species and can
change as the medium cools down as it evolves.  In this paper, we will take all
three to be of the same order. Therefore, a non-relativistic treatment is
still applicable. However, the integration of modes from $1/r$ to $E_b$
includes thermal effects.

Here we would like to point out that further assuming scale separations between
$E_b$, $m_D$, and $T$ can allow us to write simpler EFTs assuming specific
choices of these hierarchies. These have been investigated in detail in a 
series of papers~\cite{brambilla20081,brambilla20101,brambilla20111,brambilla20131}. Our goal in this paper is to avoid assuming
a clear separation between the three scales. In specific regimes where the
separations exist, our results will clearly reduce to results
from~\cite{brambilla20101,brambilla20131}. However, we shall see that in a wide range of
parameters, physics lies in an intermediate regime where clear separations do
not exist. 

To do this we use the full perturbative form of the gluon spectral function. We
include contributions from the transverse and longitudinal gluons both in the
Landau damping (LD) regime and in the space-like regime where gluo-dissociation
occurs.  This gives a clear framework to include both processes in a unified
language and allows us to compare the contributions to decay from the various
process. This is the first time the full gluonic spectral function applicable
in both kinematic regimes has been used to compute the total decay rates. In
our calculation, the singlet wavefunction is approximated to be the
instantaneous eigenstate of a lattice inspired thermal potential and hence
incorporates screening. For the octet state, the spectrum is fixed by the
constraint that at large $r$ the real part of the octet potential approaches
the real part of the singlet potential. For the wavefunction, we systematically
compare two limiting cases. One where the screening is strong  that the
potential is flat in $r$ and the other where the screening is very weak. These
can be seen as limiting cases of the physical situation where the screening
length is comparable to that in the singlet channel~\cite{Bala:2020tdt}. 

The comparison between $E_b$ and $T$ is shown in FIG. \ref{be} which clarifies
that these two scales are close to each other. The consequence of this is shown
in FIG.~\ref{fig:comp} where we show for the $\Upsilon(1{\rm{S}})$ state
gluo-dissociation dominates in a wide temperature region of interest.
FIG.~\ref{fig:comp2s} shows that for the $\Upsilon(2{\rm{S}})$ state also both
contributions are comparable. 

Finally, we find that (FIG. \ref{fig:kcomp}) the imaginary potential
over-predicts the contribution from LD substantially.  

The plan of the paper is as follows. In Sec.~\ref{sec:Formalism} we will review
the formalism and highlight the assumptions and approximations involved in our
method. In Sec.~\ref{sec:diffusion}, we discuss the connection between $EE$ correlator and the momentum diffusion coefficient of heavy quark, particularly, in the static limit. In Sec.~\ref{sec:decay}, we discuss the implementation of the real part of the singlet potential to obtain the singlet wave function at a given T. We also discuss the two extreme cases of complete screening and no screening for octet interactions. Finally, in Sec.~\ref{sec:results}, we discuss our results followed by the conclusion and future directions in Sec.~\ref{sec:conclusion}.  

\section{Formalism~\label{sec:Formalism}}
pNRQCD~\cite{brambilla20001} is an EFT for bound states of
quarkonia. In vacuum, it relies on the hierarchy of scales $M\gg \frac{1}{r}\gg E_b$.
The scale separation $M\gg \frac{1}{r}$ ensures that the $Q$ and $\bar{Q}$
are non-relativistic, and $\frac{1}{r}\gg E_b$ means that the interactions
between $Q$ and $\bar{Q}$ (at leading order in $1/M$) can be written as
potentials. 

One can think of it as a two-step process where relativistic dynamics of $Q$
and $\bar{Q}$ are integrated out first, to obtain NRQCD at scales
$1/r$~\cite{bodwin19941}.  If $\frac{1}{r}\gg T, m_D$, the energies corresponding
to the thermal scales is much smaller than the relative momentum between
$Q\bar{Q}$ ($\sim 1/r$) then NRQCD at this scale is unaffected by $T, m_D$, and
hence this theory is the same as the theory in
vacuum~\cite{bodwin19941}.  

The pNRQCD lagrangian is obtained by integrating out modes from $1/r$ to $E_b$.
The structure of the EFT is governed only by the symmetries and the particle
content of the theory. In the rest frame of $Q\bar{Q}$ in vacuum, this theory
has been extensively studied~\cite{brambilla20001}. 

While the medium introduces a new vector $u^\mu$ associated with the medium rest frame which can lead to additional operators in the
lagrangian~\cite{brambilla20111,brambilla20131}, we only consider the case where the quarkonium is (nearly) at rest in the medium, and hence the form of the lagrangian is unchanged from that in vacuum. The lagrangian is of the form
\begin{equation}
	\begin{split}
		\mathcal{L} &= S^\dagger\bigg(i\partial_t + \frac{\nabla^2}{M} -V_s(r)\bigg)S\\
		&+ O^\dagger\bigg(i\partial_t + \frac{\nabla^2}{M} - V_o(r)\bigg)O\\
		&+gV_A(r) [S^\dagger r\cdot E O +O^\dagger r \cdot E S]\\
		&+gV_B(r) [O^\dagger r\cdot E O + O^\dagger O r\cdot
		E]+\cdot\cdot\;.~\label{eq:LpNRQCD} 
	\end{split}
\end{equation}
Here $S (O)$ is the singlet (octet) wavefunction in the relative coordinate between
$Q$ and $\bar{Q}$. $V_s(V_o)$ is the $Q\bar{Q}$ potential in the singlet (octet) 
channel. $M/2$ is the reduced mass. 

The lagrangian (Eq.~\ref{eq:LpNRQCD}) is obtained by systematically performing
a multipole expansion which encodes the factorization of wavelengths of the
order of $1/E_b$ compared to the short distance $r$. The dots represent higher
order terms in this expansion. 

The low energy coefficients (LEC's) $V_A(r)$, $V_B(r)$ are $1$ at leading order
in perturbation theory and are expected to be close to $1$ at a short distance. In our paper, we will take them to be $1$. The other input to the theory are the potentials, $V_s(r)$ and $V_o(r)$. If $E_b\sim T, m_D$, then the integration of
modes from $1/r$ to $E_b$ is affected by the medium and hence the functional forms of $V_s(r)$ and $V_o(r)$ is different from their forms in vacuum.  If the $Q$ and $\bar{Q}$ can be treated as static (for example if their mass is so high that their kinetic energy can be ignored), then one can run the integration of modes all the way to zero energy. It is well known that in this limit $V_s$ and $V_o$ are complex~\cite{laine20071,brambilla20081}. The real and imaginary
parts of the static potentials have been calculated in weak coupling
limit~\cite{laine20071,brambilla20081,akamatsu20131}. Moreover, we
also expect that non-perturbative contributions to the potential are
substantial especially for the excited states of bottomonia, because while
$1/r$ is large compared to $\Lambda_{QCD}$ the hierarchy is not very strong. 
Additionally, neither $E_b$ nor $T$ are much larger than $\Lambda_{QCD}$ and
hence the medium itself at this scale is strongly coupled.

Both the real and imaginary parts of
$V_s$~\cite{rothkopf20111,Burnier:2014ssa,petreczky20121,Bala:2019cqu,Bala:2021fkm}
and $V_o$~\cite{Bala:2020tdt} have been computed
non-perturbatively on the lattice. 

Let us note that the static calculation gives the $Q\bar{Q}$ potential under
the assumption that $E_b$ is the smallest scale in the problem and the kinetic
energies of the $Q$ and $\bar{Q}$ are negligible. However, if $E_b$ and $T$ are
comparable, one needs to go beyond the static approximation. In this case the
thermal losses can not be captured by an imaginary potential. 

In this work we assume that the real part of the potential at scale $E_b$ is
captured by the static value. Thus, we take the real part of the potentials to
be in the range considered by lattice inspired potentials~\cite{Islam:2020bnp}.
On the other hand to estimate losses due to thermal process, we compute the imaginary part of the singlet self
energy diagram in the multipole expansion.  The key assumption here is that in
the pNRQCD lagrangian~(Eq.\ref{eq:LpNRQCD}) at scale $\sim E_b$, the imaginary part of
the potential is small and losses predominantly arise from dynamics at scales
$\sim E_b$. The advantage of this approach is that it captures finite
frequency effects of thermal excitation due to the medium.
However, this approach misses finite frequency effects in the real
parts of the self-energy. These corrections will change the energy of
the singlet states and similarly for the octet state and hence change the
binding energy of the singlet states. However, the effect of this contribution
on the decay rate of the singlet state is higher order in the multipole 
expansion ($r^4$ instead of $r^2$) and can be safely ignored in our
calculation.

Finally, we assume that the octet state, once formed, decoheres rapidly and can no longer lead to a reformation of the singlet state. In quantum
calculations~\cite{brambilla20171,akamatsu20191,Sharma:2019xum} these processes can be taken into account but this is beyond the scope of our paper.

\begin{figure}[h]
\centering 
\includegraphics[width=.45\textwidth,origin=c,angle=0]{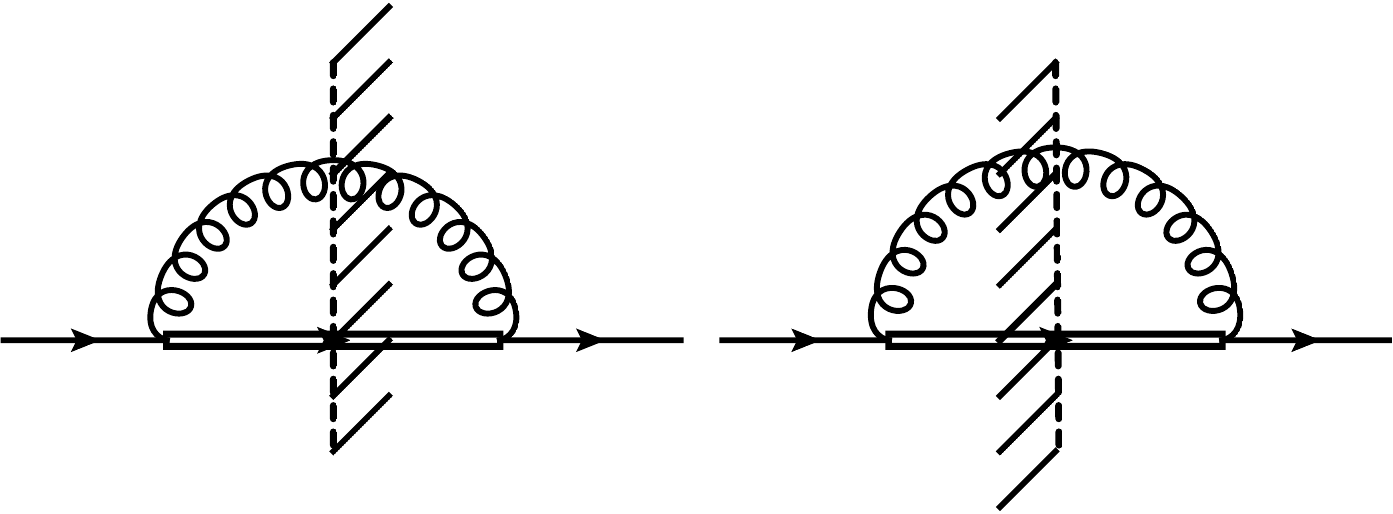}
\caption{\label{diagram}Cut diagrams contributing to the decay width. Single solid
line is for singlet and double lines for octet. The gluon line corresponds to a
dressed gluon.} \label{fig:singletself}
\end{figure}

With this setup, let us start from a singlet state. The dissociation is given
by the imaginary part of the singlet self-energy correction. The corresponding
diagram is shown in FIG.~\ref{fig:singletself}.  Here, the gluon line is
resummed and gets contributions both from LD which arises from the
imaginary part of the gluon self-energy and pole of the gluon propagator.  In
FIG.\ref{fig:singletself}, gluon momentum ($k_0$,${k}$) is directed inward at
the first vertex and octet momentum ($q_0$,${q}$) is directed outward from the
same vertex. The incoming momentum of the singlet is $p_\mu=q_\mu-k_\mu$.

In order to calculate the imaginary part of the singlet self-energy, we follow
the cutting rules at the finite temperature are given in
Refs.~\cite{Kobes:1990ua,Kobes:1986za}. There are two cut diagrams as shown in
FIG.~\ref{fig:singletself} . Following the cut rules
and implementing appropriate propagator for each cut diagram, the imaginary
part of the self-energy reads as
\begin{eqnarray}
	&\Im \Sigma_{11}(p_0,p,r)=\frac{g^2C_F}{6} r_i\bigg(\int \frac{d^4k}{(2\pi)^4}\bigg\{\rho_{o}\theta(q_0)(\theta(-k_0)\nonumber\\
	&\qquad+f(|k_0|))[k_0^2 \rho_{jj}(k_0,k)+k_j^2 \rho_{00}(k_0,k)]\bigg\}\nonumber\\
	&+\int\frac{d^4k}{(2\pi)^4}\bigg\{\rho_o \theta(-q_0)(\theta(k_0)+f(|k_0|))\nonumber\\
	&\times[k_0^2 \rho_{jj}(k_0,k)+k_j^2  \rho_{00}(k_0,k)] \bigg\}\bigg)r_i.
	\label{eq:rhoij0}
\end{eqnarray}
To avoid this lengthy expression, from here onwards, we use the following shorthand notation for the above equation
\begin{equation}
	\Im \Sigma_{11}(p_0,p,r)=r_i \hat{\cal{O}}(p_0,p,r) r_i.
	\label{eq:rhoij}
\end{equation}
In Eq.~\ref{eq:rhoij0}, $C_F$ ($=4/3$) is color factor. $\rho_{00}(k_0,k)$
and $\rho_{jj}(k_0,k)$ are gluon spectral functions that are discussed in the next section. $p_\mu=(p_0,{\bf{0}})$ is the four-momentum of the incoming singlet state and $\rho_o$ is the tree level $Q\bar{Q}$ spectral function in the octet channel which can be obtained from the octet propagator (see Eq.~\ref{eq:LpNRQCD})
\begin{equation}
	G(q_0) = \frac{1}{q_0-\nabla^2/M-V_o}.
\end{equation}
Hence, 
\begin{equation}
	\begin{split}
		\rho_o&=2\pi  \delta(k_0+p_0-\hat{q}_0),\;\;{\rm{where}}\\
		\hat{q}_0&=V_o+\frac{\nabla^2}{M}\;.
	\end{split}
\end{equation}

To proceed further we need information about the temporal and spatial gluonic
spectral functions, $\rho_{00}$ and $\rho_{ii}$, which we discuss below.

\subsection{Gluon-polarization tensor}
In this section, we review the well known expressions for the gluon
polarization tensor~\cite{kapusta_gale_2006} that are essential
inputs to the evaluation of the imaginary part of quarkonium self-energy.

The general form of the gluon self-energy is given as
\begin{equation}
\Pi_{\mu \nu}(k_0,k)=P^L_{\mu \nu} \Pi_L(k_0,k)+P^T_{\mu \nu} \Pi_T(k_0,k),
\end{equation}
where $P^L_{\mu \nu}(P^T_{\mu \nu})$ are longitudinal (transverse) projection operators and $\Pi_L (\Pi_T)$ are component of  the gluon self-energy along these directions. In order to evaluate quarkonium decay width using Eq.~\ref{eq:rhoij}, we need the imaginary part of the gluon propagator. This may come from the pole of the propagator in the region of phase space where the imaginary part of the gluon self-energy is zero, and from the region where the imaginary part of the gluon self-energy is finite. The pole contribution is non-vanishing in the limit $k_0>k$, and the latter contribution that requires the real and imaginary parts of the gluon self-energy is finite when $k_0< k$~\cite{kapusta_gale_2006}. Below we discuss various components of the gluon self-energy. 

Let us first consider the regime $k_0<k$ (space-like). This we call the Landau
damping regime.

The gluon loop contribution to the imaginary part of the longitudinal component
of the gluon self-energy is given as
\begin{eqnarray}
\Im \Pi^g_{L}(k_0,k)&=& \frac{g^2 N}{4 \pi k}  \int_{\frac{k+k_0}{2}}^{\infty} dq \,q^2\,\bigg(2+\frac{k^4}{4 q^4}-\frac{k^2}{q^2}\bigg)\nonumber\\
&\times &(f(q-k_0)-f(q))\,\theta(k-k_0),
\label{eq:imag}
\end{eqnarray}
where $N=3$ and  $f(q)$ is Bose-Einstein distribution function.  Let us note
that with an expansion in $k_0/T$ in the distribution function and by taking
the limit $k\ll q$, Eq.\ref{eq:imag} goes to its hard thermal loop (HTL)
counterpart. 

Similarly, The quark loop contribution with $N_f$ (light) quark flavors to the imaginary part of the  longitudinal component of the gluon self-energy is given as
\begin{eqnarray} 
\Im \Pi_L^f(k_0,k)&=&\frac{ g^2 N_f}{2 \pi k}\int dq \,\bigg(q^2-\frac{k^2}{4}\bigg)\,(\tilde{f}(q-k_0)\nonumber\\
&-&\tilde{f}(q))\,\theta(k-k_0),
\label{eq:imagselfq}
\end{eqnarray}
where $\tilde{f}(q)$ is Fermi-Dirac distribution function. The total imaginary
part of the longitudinal gluon self-energy can be obtained by summing
Eqs.~\ref{eq:imag} and \ref{eq:imagselfq}. 

   The real part of the longitudinal component of the self-energy is
\begin{equation}
\Re\Pi_{L}(k_0,k)=m_D^2\bigg(1-\frac{k_0}{2 k}\log\bigg|\frac{k+k_0}{k-k_0}\bigg|\bigg),
\label{eq:reall}
\end{equation} where $m_D^2=\frac{g^2 T^2}{3}\big(N+\frac{N_f}{2}\big)$. In
obtaining Eq.~\ref{eq:reall}, we have dropped terms of the order of
$k_0/T,k/T$. These terms are important when $k_0,k\gtrsim T$ but we drop these
terms because of the following reason. The exact forms of these higher order
terms depend on the gauge (for eg. see ~\cite{kapusta_gale_2006}) while
Eq.~\ref{eq:reall} is the HTL form and is gauge
invariant~\cite{Braaten:1990it}. This expression is valid for both $k_0>k$ and
$k_0<k$.  Moreover, from Eq.~\ref{eq:rhoij} it is clear that the contribution
to $\Im\Sigma$ from  $k_0,k\gg T$ is exponentially suppressed and hence making 
this approximation will not cause a significant error in our result. 

Similarly, for the transverse gluon, the imaginary contributions to the gluon self-energy are,
\begin{eqnarray}
\!\!\!\!\Im \Pi^g_{T}(k_0,k)&=&\frac{g^2 N}{2 \pi k } \int_{\frac{k+k_0}{2}}^{\infty} dq\, \bigg[q^2\bigg(1-\frac{k^2}{2 q^2}\bigg)^2-\frac{k^2}{4}\nonumber\\
&\times&\!\!\bigg(\!2\!-\!\frac{k^2}{2 q^2}\!\bigg)^2\bigg]\!(f(q\!-\!k_0)\!-\!f(q))\theta(k\!-\!k_0),\nonumber\\ 
\label{eq:imagt}
\end{eqnarray}
and, 
\begin{eqnarray}
\Im \Pi_T^f(k_0,k)&=&\frac{ g^2 N_f}{8 \pi k}\int_{\frac{k+k_0}{2}}^{\infty} dq\, \bigg(2q^2+\frac{k^2}{2}\bigg)\nonumber\\
\!&\times&(\tilde{f}(q-k_0)-\tilde{f}(q))\,\theta(k-k_0).
\end{eqnarray}

The real part of the transverse component of the gluon self-energy is
\begin{equation}
\!\!\!\Re\Pi_{T}(k_0,k)= \frac{m_D^2}{2}\bigg(\frac{k_0^2}{k^2}-\frac{k_0(k_0^2-k^2)}{2k^3}
\log\bigg|\frac{k+k_0}{k-k_0}\bigg|\bigg).
\label{eq:realt}
\end{equation}

Let us now consider the regime $k_0>k$ (time-like). This we call the pole
regime. In this regime, the imaginary part of $\Pi_L$ and $\Pi_T$ are zero and
the real parts are as above. At order $g^3$ the widths of these modes are
finite~\cite{kapusta_gale_2006,Bellac:2011kqa} but we ignore this in our
calculation.

Now we can calculate the  gluon spectral function which goes in
Eq.~\ref{eq:rhoij}. Below we discuss it for both time-like and space-like
gluons.

\subsection{Gluon spectral functions}

The general form of the gluon spectral function in a medium reads as
\begin{equation}
\rho_{\mu \nu}(k_0,k)={{P}}^L_{\mu \nu} \rho_L(k_0,k)+{{P}}^T_{\mu \nu} \rho_T(k_0,k).
\label{eq:rhocut}
\end{equation}
Here $\rho_L(k_0,k)=D^R_{L}(k_0,k)-D^A_{L}(k_0,k)$ is the longitudinal
component of the spectral function and $D^{R(A)}_L$ is longitudinal component
of resummed retarded (advanced) gluon propagator. Similarly, one can obtain the transverse component of the spectral function ($\rho_T$) by using
$\rho_T(k_0,k)=D^R_{T}(k_0,k)-D^A_{T}(k_0,k)$. Below we discuss the form of these spectral functions for both $k_0>k$ as well as $k_0<k$. 

For $k_0<k$ (LD), we use the gluon self energies (shown in the previous
section) to write the resummed gluon propagator and obtain     
\begin{equation}
	\rho_{L}(k_0,k)=\frac{2 \Im \Pi_{L}(k_0,k)}{(k^2+\Re\Pi_{L}(k_0,k))^2+(\Im \Pi_{L}(k_0,k))^2},
	\label{eq:rhol}
\end{equation}
where $\Re \Pi_L (\Im \Pi_L)$ is sum of both gluon and quark contributions.

Similarly, the transverse component of the spectral function can be written as
\begin{equation}
\!\rho_{T}(k_0,k)\!=\!\frac{2 \Im \Pi_{T}(k_0,k)}{(k_0^2\!-\!k^2\!+\!\Re\Pi_{T}(k_0,k))^2\!+\!(\Im \Pi_{T}(k_0,k))^2}.
\label{eq:rhot}
\end{equation}
It is worth mentioning here that the above form of the spectral functions
reproduces the momentum diffusion coefficients obtained within the kinetic
theory framework in Ref.~\cite{moore20051}.

For $k_0>k$ the gluon propagator is simply a pole. The quarkonium dissociation
in this regime is due to the absorption of a gluon from  thermal medium. This
process is known as gluo-dissociation in the literature. In the limit $k_0\gg
T$, the spectral function is given by the imaginary part of free gluon retarded
propagator and gluo-dissociation in this case has been studied in
Refs.~\cite{brambilla20111,Sharma:2019xum}  However, for realistic
situations one needs to take full resummed propagator. Thus, similar to
Eq.~\ref{eq:rhocut} the general form of the spectral function reads as
\begin{equation}
\rho^p_{\mu \nu}(k_0,k)={\cal{P}}^L_{\mu \nu} \rho_L^p(k_0,k) +{\cal{P}}^T_{\mu \nu}\rho_T^p(k_0,k),
\label{eq:rhopole}
\end{equation}
where $p$ stands for pole. The longitudinal spectral function in this regime 
is given by
\begin{equation}
\rho_L^p(k_0,k)=2 \pi \delta(k^2-\Re\Pi_L(k_0,k))\;.~\label{eq:plasmino}
\end{equation}
The transverse spectral function 
is given by
\begin{equation}
\rho_T^p(k_0,k)=2 \pi
\delta(k_0^2-k^2-\Re\Pi_T(k_0,k)).~\label{eq:plasmon}
\end{equation}

The imaginary part of $\Sigma_{11}$ gets contribution from both
Eqs.~\ref{eq:rhocut} and \ref{eq:rhopole}. While in the low frequency limit,
LD gives the dominant contribution, pole contributions are
significantly large in the intermediate and high frequency limit. The overall
pole contribution merges with their free spectral function counterpart at
an asymptotically large frequency. This we show in FIG.~\ref{fig:spectral}.

\section{Connection with the momentum diffusion coefficient~\label{sec:diffusion}}
In this section, we relate Eq.~\ref{eq:rhoij} with the standard definition of
the momentum diffusion coefficient in terms of the electric field correlator which is 
given as~\cite{solana20061,huot20081}
\begin{equation}
	\!\!\!\!\kappa=\!\frac{g^2}{3 N}\!\!\int_{-\infty}^{\infty}\!\!\!\!\!dt\, \text{Tr}\langle U\!(-\infty,\! t)E_i(t)U\!(t,\!0)E_{i}(0)U\!(0,\!-\infty)\rangle, 
	\label{eq:ee}
\end{equation}
where $U$ is the Wilson line in the fundamental representation, $E_i=\partial_iA_0-\partial_0 A_i$ is the color electric field and  trace over color degrees of freedom. In Eq.~\ref{eq:ee}, the infinite integration limit represents the zero frequency limit of the correlator.  For the leading order results, one needs to replace the Wilson lines by  identity (i.e., $U=\mathds{1}$) to obtain
\begin{eqnarray}
	\kappa&=&\frac{g^2C_F}{3}\lim_{k_0\rightarrow 0 }\int \dbar^3k\, k^2\, \langle A_0(k_0,k)A_0(0,0)\rangle \nonumber\\
	&=& \frac{g^2C_F}{3}\lim_{k_0\rightarrow 0 } \int  \dbar^3k\,k^2(1+f(k_0))\rho_{L}(k_0,k),
	~\label{eq:ee1}
\end{eqnarray} 
for more details see Refs.\cite{moore20051,huot20081,Francis:2015daa}.

It is useful to compare this quantity (Eq.~\ref{eq:ee1}) to the expression
Eq.~\ref{eq:rhoij}. Imposing the condition $q_0>0$ in Eq.~\ref{eq:rhoij} and  performing energy integration in Eq.~\ref{eq:rhoij0} using the energy delta function we rewrite the imaginary part of the singlet self-energy as
\begin{equation}
	\begin{split}
		\Im \Sigma_{11}(k_0)=
		\frac{g^2r^2C_F}{6} \times& \\
		\int\dbar^3 k f(k_0)
		&\Bigl[k_0^2 \rho_{ii}(k_0,k) 
		+k^2 \rho_{00}(k_0,k)\Bigr],
	\end{split}
	\label{eq:rhoij4}
\end{equation}
where   $k_0=p_0-q_0$. For future use, we define a quantity
\begin{equation}
	\tilde{\kappa}(k_0)=\frac{2(\Im\Sigma_{11}(k_0)|_{\text{pole}}+\Im\Sigma_{11}(k_0)|_{\text{LD}})}{r^2}.
	\label{eq:kappa}
\end{equation}
A single heavy quark, traversing through the thermal medium, gets uncorrelated
random kicks from the medium constituents that give rise to $\kappa$. However,
for quarkonium bound state, not only scattering but also absorption of thermal
gluons  contribute to the dissociation. The latter process is kinematically forbidden for a single heavy quark. Therefore, in the frequency regime where
gluo-dissociation dominates $\tilde{\kappa}$ is not the same as $\kappa$.
However, in the static limit where dissociation via scattering (i.e., LD) is dominant, the two coefficients defined in Eqs.~\ref{eq:ee} and
\ref{eq:kappa} seem identical, at the leading order. We have checked that the LD
part of $\tilde{\kappa}$ agrees with the one obtained in Ref.~\cite{moore20051}.

We note that at higher order there is no reason for these two coefficients to be identical. The reason is that for quarkonium, chromo-electric field correlator is defined with Wilson lines in the adjoint
representation~\cite{Eller:2019spw}. On the other hand for a single heavy
quark, Wilson lines are in fundamental representation.

It is useful to note here that in Eq.~\ref{eq:kappa} if one makes  $k_0,k\ll
T$ approximation {\it{before}} integrating over $k$, the integrand is of the
HTL form and is ultraviolet (UV) divergent. This is due to the fact that the applicability of
the  HTL resummation is restricted to the low frequency limit. This divergence
can be cured by either using a cutoff $k\sim m_D$ or by adding the UV
contribution to the integral  carefully~\cite{brambilla20131}. However, the
contribution from Eqs.\ref{eq:rhol} and \ref{eq:rhot} vanishes in the high
frequency limit. Therefore, if we use the imaginary part of the longitudinal
self-energy given by Eqs.~\ref{eq:imag} and \ref{eq:imagselfq} and evaluate
$\tilde{\kappa}$, the integral is convergent and can be computed numerically
which we do next. 

 \begin{figure}[h]
	\centering 
	\includegraphics[width=.43\textwidth,origin=c,angle=0]{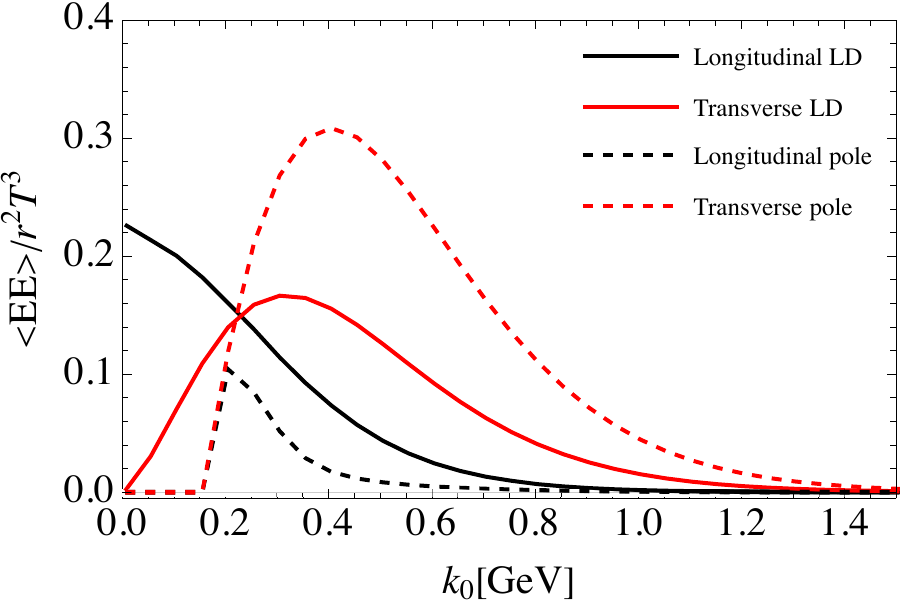}
	\caption{\label{spec1} Various contributions to scaled $\langle EE \rangle$ correlator as a function of frequency. Here we take $T=0.25$ GeV and $m_D=0.5$ GeV.}
	\label{fig:eecorr}
\end{figure}

In FIG.~\ref{fig:eecorr}, we plot the electric field correlator arising from
various contributions that appear in the evaluation of singlet self-energy
diagram in FIG.~\ref{fig:singletself} as a function of frequency. Here we take
$T=0.25$ GeV and Debye mass $m_D=0.5$ GeV. The black curves are for the
longitudinal gluon with a solid line for LD and a dashed one for pole
contributions. The red curves are for transverse gluon where solid and dashed
lines are for LD and pole contributions, respectively. As anticipated, in the
small frequency limit the dominant contribution comes from longitudinal gluon
Landau damping. In this limit, other contributions are either zero or very
small. Pole contributions switch on at a somewhat larger frequency, i.e.,
$k_0\sim 0.2$ GeV. Moreover, the transverse gluon pole contribution is larger (in
magnitude) compared to the longitudinal one. Finally,  at high frequency,
transverse pole contribution dominates and eventually approaches the corresponding free limit. 

\begin{figure}[h]
	\centering 
	\includegraphics[width=.43\textwidth,origin=c,angle=0]{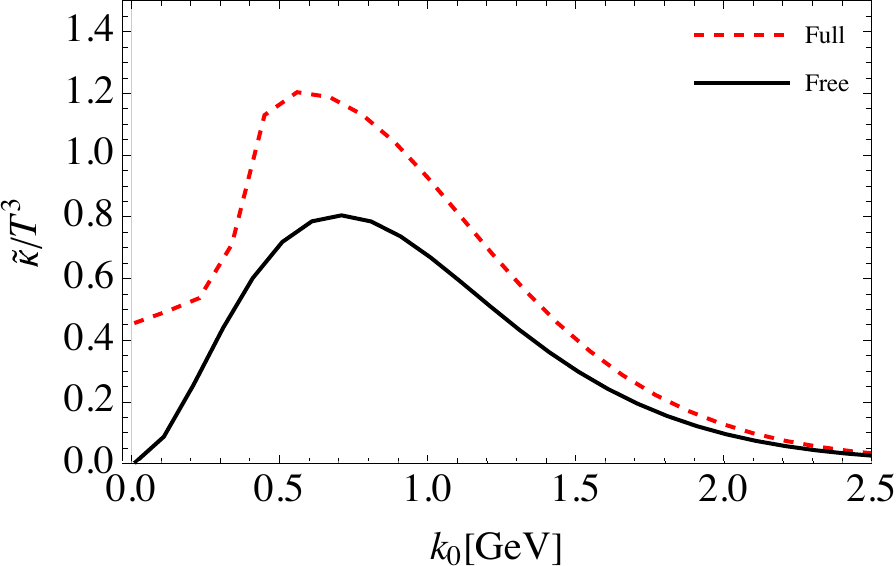}
	\caption{$\tilde{\kappa}/T^3$ as a function of frequency for constant coupling $g=2$, $T=0.25$ GeV and $N_f=3$. Black (solid) line is the free limit and the dashed (red) is resummed one.}
	\label{fig:spectral}
\end{figure}

In FIG.~\ref{fig:spectral}, we have plotted $\tilde{\kappa}/T^3$ as a function
of $k_0$. The red (dashed)
curve here gets contribution from both $k_0<k$ as well as $k_0>k$ phase space
regions.  In the static limit, i.e., $k_0\approx
0$, we have checked that $\tilde{\kappa}/T^3$ agrees with that in 
Ref.\cite{moore20051}. The black (solid) line is in the free limit which  as expected is zero at
zero frequency.

It is well known that the perturbative result for
$\kappa$ is too low by roughly a factor of $5-10$ than the non-perturbative
value~\cite{datta20121,laine20151,Banerjee:2022gen,Brambilla:2022xbd}. For example, recent 
lattice results for $\kappa/T^3$ are estimated as~\cite{Brambilla:2019oaa}
\begin{equation}\nonumber 
	1.99 < \frac{\kappa}{T^3}< 2.69  \hspace{1cm} \text{for}  \hspace{0.5cm} T=1.5 T_c,
\end{equation}
\begin{equation}\nonumber 
	1.05 < \frac{\kappa}{T^3}< 2.26  \hspace{1cm} \text{for}  \hspace{0.5cm} T=3 T_c.
\end{equation}
For finite $k_0$ there are no lattice results available in the literature.
Naively, we expect them to be different from the perturbative estimates.
Therefore, we expect that our predictions for $R_{AA}$ are
underestimated. However, our results capture the qualitative features of
relative contributions of pole and LD in the range of temperatures available in
HIC. Motivated by lattice QCD calculations of $m_D$, we choose
$g=2$~\cite{Kaczmarek:2005ui,Islam:2020bnp}.  

In the intermediate frequency regime, the peak structure in $\tilde{\kappa}/T^3$ is from the transverse pole
contribution. Finally, in the high frequency limit, it merges with its free
limit counterpart.   

\section{Decay width~\label{sec:decay}}
At any given temperature $T$, the decay width of a singlet state $|\phi\rangle$ at leading order is given by
\begin{equation}
\Gamma =2\, \langle \phi|\Im \Sigma_{11}|\phi\rangle, 
\label{eq:decayw}
\end{equation}
where $\Im \Sigma_{11}$ is given in Eq.~\ref{eq:rhoij}. Inserting a complete
set of octet states $|o\rangle\langle o|$ at the right bracket of
Eq.~\ref{eq:rhoij}, one obtains
\begin{eqnarray}
\Gamma &=&2\,\sum_{o}\langle \phi|r\hat{\cal{O}}(p_0,p, r)|o\rangle \langle o|r|\phi\rangle \nonumber\\
&=& 2\,{\cal{O}}(p_0,p)\sum_{o} \langle \phi | r |o\rangle\langle o|r|\phi\rangle\;,
\label{eq:decay}
\end{eqnarray}
where summation is over all octet states allowed by the selection rule. In
operator form ${\cal{O}}(p_0,p)$ is same as defined in Eq.~\ref{eq:rhoij} and
now $p_0=q_0-k_0$. 

Let us note that the octet state lies in the continuum. $q_0$ is the energy of the octet state,
which is given by,
\begin{equation}
q_0|o\rangle=\bigg(V_o+\frac{\hat{q}^2}{M}\bigg)|o\rangle.  
\label{eq:energy_cons}
\end{equation}

One point to note is that momentum conservation implies that the center of mass
momentum of the octet state is $-{{k}}$. This implies that the energy of the
$|o\rangle$  state has an additional contribution $k^2/(4M)$ which should be
added to the right hand side of Eq.~\ref{eq:energy_cons}. The value of $k$ is
governed by $T$ since it is the region in $k$ space where the gluon spectral
function multiplied by the Bose-Einstein distribution function is not exponentially
suppressed. In the hierarchy we are working, $E_b$ and $T$ are both  small
scales compared to $M$ and hence quantities of the order of $T^2/(4M)$ are suppressed by an
extra power of $M$ compared to the right hand side of Eq.~\ref{eq:energy_cons} and
hence can be safely dropped.

\subsection{Modelling the singlet state}
To complete the evaluation of Eq.~\ref{eq:decay}, we need the functional form of
the singlet state $|\phi\rangle$ as well as octet state $|o\rangle$. Below we discuss the prescriptions to obtain these wave functions. 

For a state created in vacuum and ``dropped'' into the QGP, a natural choice
for $|\phi\rangle$ is the wavefunction in vacuum.  Further, if
the thermal effects are weak then $|o\rangle$ can be taken to as octet states
in vacuum. If the initial formation of quarkonia is not affected by the medium
(for example if the formation of the quarkonium states occurs on a time scale
much shorter than the formation of the QGP) then this is a well motivated model
for $|\phi\rangle$ and $|o\rangle$. This picture has been  previously
used for phenomenology ~\cite{Park:2007zza,Sharma:2012dy}. 

At the LHC and the RHIC, the formation time of  the QGP is a fraction of a
fm/c and is not substantially larger than the formation time of quarkonia, of
the order of $1/E_b$. One can expect the formation dynamics of quarkonia to be
substantially affected by the medium.

One natural way to include these effects is to start the evolution from a
narrow initial $Q\bar{Q}$ state of width $\sim M$ and follow its quantum
evolution from very early time~\cite{Islam:2020bnp,brambilla20171}. In this
paper, we do not study the quantum dynamics and this analysis is beyond the
scope of the paper. If dissociation can be modelled by the imaginary part of
the potential (i.e. in the $E_b\ll T$ regime) another possible approach is to
assume that the evolution dynamics is slow (adiabatic approximation) and the
quarkonium state is initially formed in the eigenstate of the complex potential
and at each instant the quarkonium state is in the eigenstate of the complex
potential~\cite{Strickland:2011mw,strickland20112}. In this paper, dissociation is
calculated using Eq.~\ref{eq:rhoij} which can not be captured by a complex
potential and hence the adiabatic method is not applicable.  We model the
effect of the medium on the formation of quarkonia by making the maximal
approximation that the initial state and subsequent to formation is determined
by the real part of the instantaneous thermal potential \cite{zhao20121}. 

More concretely, for the singlet states wavefunction, we use the eigenstates
\begin{equation}
p_0|\phi\rangle=\bigg(\frac{p^2}{M}+V_s(r, T)\bigg)|\phi\rangle,
\label{eq:singlet}
\end{equation} 
where $V_s(r, T)$ is the real part of the thermal potential. Here we have subtracted the rest energy from all the $Q\bar{Q}$ states. Similarly,
$|o\rangle$ is given by Eq.~\ref{eq:energy_cons} with $V_o$ given by the real part
of the octet potential in the thermal medium. In summary, to calculate
$|\phi\rangle$ and $|o\rangle$ we need the real parts of the potentials $V_s$,
$V_o$. 

\begin{table}
\begin{tabular}{|c c c c c c|}
\hline 
 &$\Upsilon$(1S) & $\Upsilon$(2S) & $\chi_b$(1P) & $\Upsilon$(3S) & $\chi_b$(2P)  \\
 \hline 
$M_M$ & 9.46 & 10.0 & 9.88 & 10.36 & 10.25 \\
$E_b$ & 1.20 & 0.66 & 0.78 & 0.30 & 0.41 \\
$\langle r^2 \rangle $ & 1.42 & 6.58 & 4.20 & 13.68 & 10.60\\
\hline 
\end{tabular}
\caption{Binding energies, mass and $\langle r^2 \rangle$ of various bounds
state at $T=0$ using Eq.~\ref{eq:pot}. All dimensions are in GeV.~\label{table:BE}}
\end{table}

For the singlet potential we use the lattice inspired potential which is given
by Refs.\cite{Dumitru:2009ni,Islam:2020bnp}
\begin{equation}
V_s(r,T)\!=\!-\frac{a}{r}(1+m_Dr)e^{\!-m_Dr}\!+\frac{2\sigma}{m_D}(1\!-e^{\!-m_D r})\!-\sigma re^{\!-m_D r}\!\!.
\label{eq:pot}
\end{equation}

The effective coupling $a=0.409$ and the string tension $\sigma=0.21$ GeV$^2$
are fixed from the vacuum masses and binding energies (see
TABLE~\ref{table:BE}) with bottom mass $M=4.7$ GeV. Here we take $m_D=0$ for
obtaining the vacuum spectrum. 

For finite $T$ we keep $a$ and $\sigma$ the same as in $T=0$ and
$m_D=\sqrt{(1+N_f/6)}gT$.  For $g=2$ Eq.\ref{eq:pot} gives potentials
consistent with those used for bottomonium phenomenology with lattice based
potentials~\cite{krouppa20171,Burnier:2016mxc}. The $Q\bar{Q}$ potential in the medium is screened, as a
result of which $E_b$ becomes smaller with increasing
temperature. At sufficiently high temperature the bound state is
dissolved~\cite{matsui19861}. It is worth mentioning that for the 1S state, the wavefunction does
not depend on the temperature of the medium up to $T\sim 480$ MeV and it
remains approximately the same as that of vacuum Coulombic
state while the excited states dissolve earlier~\cite{ymocs20071}.  
 
\begin{figure}[h]
\centering 
\includegraphics[width=.43\textwidth,origin=c,angle=0]{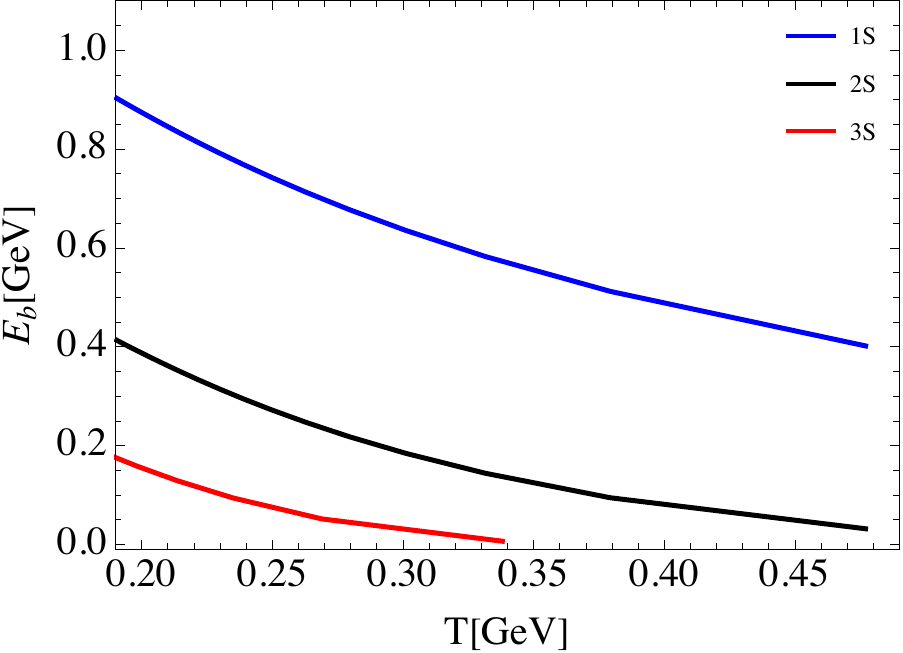}
\caption{\label{be}Binding energies for $\Upsilon$(1S), $\Upsilon$(2S) and $\Upsilon$(3S) states as a function of temperature. Here we take constant coupling $g=2$.}
\end{figure}

 In FIG.~\ref{be}, we plot the binding energy of $\Upsilon$(1S), $\Upsilon$(2S)
and $\Upsilon$(3S) states as a function of medium temperature. For a given
potential, $E_b$ is given by $E_b=2 M-M_M+V_{\infty}$, where $M$ is bottom
current mass, $M_M$ is bound state mass and $V_{\infty}$ is the asymptotic value
of real part of singlet potential. The key point we want to highlight in
FIG.~\ref{be} is that for the temperature range relevant for HICs, the
hierarchies $E_b\gg T$ or $E_b\ll T$ may not be satisfied, at least for 1S and
2S states.  It is worth mentioning that $E_b$ obtained here agrees with the one in Ref.~\cite{Du:2017hss}.
On the other hand for higher states, binding energy approaches zero around
this temperature and $E_b\ll T$. 
 
A consequence of our model is that we can not address the observed
phenomenology of the 3S state~\cite{CMS:2022rna} as for the central bins $R_{AA}$ for
3S state (in our model) is zero. The key dynamics missing from the classical
model are (1) quantum formation dynamics and (2) processes that allow for reformation
of bound states  which are important for capturing 3S dynamics. 

 \subsection{Modelling the octet state}
The octet states are also affected by the thermal medium. At short distances we
know that the potential is repulsive Coulombic. In perturbation theory both the
real and the imaginary parts of the medium modified octet potential have been
computed~\cite{akamatsu20131}.  Recently, both the real and imaginary parts of
the octet potential have also been computed in pure gluonic theory on the
lattice~\cite{Bala:2020tdt}. An important outcome from these papers is that in
both perturbative and non-perturbative calculations one finds that at large $r$
the singlet potential approaches the octet potential. This is still an active
area of research but the form of the potential for $2+1$ flavor QCD is not yet
known.   

Based on these considerations, we take two limiting cases for the octet
potential. The first is when the screening is strong and one can ignore the
octet repulsion at the distance scale of interest and hence
\begin{equation}
	V_o(r,T)=V_{\infty},
\end{equation}
where $V_{\infty}=\frac{2\sigma}{m_D}$ is the asymptotic value of singlet
potential. In this case, the $|o\rangle$ wavefunctions are the same as the wavefunctions of
the free particle but the energy levels start from $V_{\infty}$. 

The other limit is where the screening is weak and  
\begin{equation}
	V_o(r,T) \rightarrow  C_F\frac{{\alpha}}{2 Nr} + V_{\infty},
	\label{eq:octet_coulomb}
\end{equation}
We expect the true physics to be between these two
limiting cases. 

Finally, the octet potential also has an imaginary piece, which corresponds to
the change of the octet state to the singlet state. In this paper, we assume
that these processes do not regenerate bound states because the octet states are much broader than the singlet state.  

For the repulsive Coulombic potential, the general form of the radial wave function  is given as~\cite{abramowitz+stegun}
\begin{equation}
R_l(\rho)=\frac{C_l}{\rho} \rho^{l+1}e^{i\rho} {}_{1}F_{1}(1+l+i\nu,2l+2,-2 i \rho),
\label{eq:radial}
\end{equation}
where ${}_{1}F_{1}$ is confluent hypergeometric function, $\rho=rp$,
$\nu=\frac{1}{8 a_0 p}$ with $a_0=\frac{2}{\tilde{ \alpha} M}$ as Bohr radius.
In this work, we take $a_0=1/1.334$~\cite{brambilla20171}.The normalization factor $C_l$ in
Eq.\ref{eq:radial} reads as
\begin{equation}
C_l=\frac{2^l e^{-\frac{\nu \pi}{2}} \sqrt{\Gamma(1+l+i\nu)\Gamma(1+l-i\nu)}}{\Gamma(2+2l)}.
\end{equation}
Let us note that for the above form of $C_l$, the wavefunction obeys the following form for normalisation
\begin{equation}
\int r^2 R_{l}(pr)R_{l}(p'r)=\frac{(2\pi)^3}{p^2}\delta(p-p').
\end{equation}
With this choice of normalisation, the decay width has the form given in Eqs.\ref{eq:LD},\ref{eq:LDT} and \ref{eq:decaypole}. With the above form of the radial wave function, the general form of the octet wave function can be written as
\begin{eqnarray}
\!\!|o\rangle& =&4 \pi R_{l}(p r)\sum_m Y_{m}^{*l}(\hat{r})Y_{m}^{*l}(\hat{p}).
\label{eq:wavefn}
\end{eqnarray} 
Here $Y_m^l$ is spherical harmonics. For the 1P state, replacing $l=1$ in the obove
equation and summing over quantum number $m$, the octet wavefunction
$|o\rangle$ reads as
\begin{eqnarray}
\!\!\!\!|o\rangle&=& \sqrt{2 \pi} \textbf{p}\cdot \textbf{r}e^{i p r} \sqrt{\frac{\nu(\nu^2+1)}{e^{2 \pi nu }-1}} {}_{1}F_{1}(2+i\nu,4,-2 i p r ).
\label{eq:wavefn1p}
\end{eqnarray} 
For $s(d)$ states, the wave function can be obtained by replacing $l=0(2)$ in
Eq.~\ref{eq:wavefn}.

In the case of no final state interaction, we take free wave function which in terms of Bessel function is given as
\begin{equation}
|o\rangle=4 \pi  j_l(p r)\sum_m Y_{m}^{*l}(\hat{r})Y_{m}^{l}(\hat{p}).
\label{eq:wavef}
\end{equation}

Below we discuss the contribution from the Landau damping and gluon absorption
in the decay width.
\subsection{Landau damping contribution} 
One way to organize the kinematic regimes which contribute to the decay width (Eq.~\ref{eq:decayw}) is 
space-like and time-like. Dissociation (of quarkonia) via scattering of the bound state with the thermal partons occurs when
the momentum of the exchanged gluon is space-like. As mentioned, this mechanism is known as LD. It is estimated by taking the resummed propagator for
the gluon line in FIG.~\ref{fig:singletself}.  The LD contribution is
dominant when  $E_b\ll m_D, T$~\cite{brambilla20081}. The overall
contribution is both from the longitudinal as well as transverse gluon.
Moreover, in the small $k_0$ limit, the longitudinal gluon contribution turns out to
be dominant. This is easily understood by noting that while both $\rho_L$ and
$\rho_T$ go as $k_0$ at small $k_0$, $\rho_T$ is multiplied by $(k_0)^2$ in
Eq.~\ref{eq:rhoij} while $\rho_L$ is multiplied by $(k_i)^2$. For $E_b\ll m_D,
T$, $k_0\ll k$, and hence the  longitudinal gluon contribution turns out to be
dominant. However, we evaluate both of them numerically and do not drop the
transverse contribution.

The explicit expression of $\Gamma_L$ for gluonic system is given (see Eq.~\ref{eq:decay}) by,
\begin{eqnarray}
\Gamma_L&=&\!\frac{C_Fg^4 N}{6 \pi}\!\!\int\!\! \dbar^3pf(k_0)\!\!\int\! \dbar^3k \frac{k\theta(k-k_0)}{(k^2+\Re\Pi_{00})^2+\Im\Pi_{00}^2}\nonumber\\
&\times&\int_{\frac{k+k_0}{2}}^{\infty}dq\,q^2\bigg(2+\frac{k^4}{4 q^4}-\frac{k^2}{q^2}\bigg)(f(q-k_0)-f(q))\nonumber\\
&\times&|\langle \phi|r|o\rangle|^2.  
\label{eq:LD}
\end{eqnarray}
For finite $N_f$, $\Re\Pi_{00}$ and $\Im\Pi_{00}$ gets contribution from the gluon as well as quark loop. Here $p$ is octet momentum and $k_0=\frac{p^2}{M}-E$ where $E$ is binding energy of the bound state. 

Assuming a strong hierarchy between $E_b$ and $T$, this can be simplified to
the one obtained in Ref.\cite{brambilla20081}.

The transverse gluon contribution arises from the transverse part of the
spectral function given in Eq.\ref{eq:rhot}. As may be noticed in
Eq.~\ref{eq:rhoij}, for finite $k_0$ this term can not be ignored. Moreover, at
significantly large $k_0$, this term can dominate over the longitudinal gluon
contribution.  This situation may arise when the hierarchy between $E_b$ and
medium temperature is not very strong. This can be observed from
FIG.~\ref{fig:eecorr} that for $k_0$ sufficiently large, the transverse
contribution to $\Im\Sigma_{11}$ can be larger than the longitudinal. It is
also clear that in the kinematic regime where the transverse LD contribution is
comparable to the longitudinal LD contribution, the binding energy and the
final state interactions can not be ignored in either.  

 Following the similar prescription as for the case of longitudinal gluon, the contribution to the decay width from the transverse gluon reads as
\begin{eqnarray}
\Gamma_T&=&\!\frac{g^4C_FN}{3\pi}\!\!\int\! \dbar^3pf(k_0)\!\!\int\! \dbar^3k\frac{k_0^2\theta(k-k_0)}{(k_0^2-k^2-\Re\Pi_{ii})^2+\Im\Pi_{ii}^2}\nonumber\\
&\times&\frac{1}{k}\int_{\frac{k+k_0}{2}}^{\infty}\! dq \, \bigg(q^2\bigg(1-\frac{k^2}{2 q^2}\bigg)^2-\frac{k^2}{4}
\bigg(2-\frac{k^2}{2 q^2}\bigg)^2\bigg)\nonumber\\
&\times&(f(q-k_0)-f(q))|\langle \phi|r|o\rangle|^2.
\label{eq:LDT}
\end{eqnarray}
Total contribution from Landau damping is sum of Eqs.~\ref{eq:LD} and \ref{eq:LDT}. 

\subsection{Pole contribution}
In the kinematic regime $E_b\gg T$, the dominant contribution to quarkonium
dissociation from singlet to unbound octet state occurs by absorbing a
time-like gluon from the thermal medium. This process is known as gluo-dissociation~\cite{brambilla20111}.  For the decay width evaluation, the
contribution to the quarkonium self-energy arises from the pole of the gluon
propagator. In the free limit, i.e., $T\gg E_b$, the medium contribution to the
singlet to octet thermal breakup appears in the thermal weight only. However,
in the intermediate temperature range, HTL effects also become important and
one needs to use resummed propagator.  In the HTL resummed propagator,  both
longitudinal and transverse gluon contribute to the imaginary part of the gluon
propagator.  After adding both of these contributions and performing $k_0$
integration using energy delta function, $\Im\Sigma_{11}$ reads as
\begin{eqnarray}
\Im \Sigma_{11}^{P}&=&\frac{ C_F g^2 r^2}{6}f(k_0)\int\! \dbar^3k \bigg(\frac{2 k_0^2 \delta(k-k_0^T)}{|{{\partial \Re\Pi_L}/{\partial k}}|_{k_0^{T}}}\nonumber\\
&+&\frac{k^2 \delta(k-k_0^L)}{|{{\partial \Re\Pi_T}/{\partial k}}|_{k_0^{L}}}\bigg),
\label{eq:pole}
\end{eqnarray}
where $k_{0}^{T}$ is solution of $k_0^2-k^2-\Re{\Pi_T}=0$ and $k_0^L$ is that of $k_0^2-\Re{\Pi_L}=0$. In the limit $T\gg E_b$, Eq.~\ref{eq:pole} can be solved analytically by making an expansion in the bosonic distribution and replacing the spectral function by the free spectral function.   Using Eq.~\ref{eq:pole}, decay width is given as
\begin{eqnarray}
\Gamma_P&=&\frac{C_F g^2 }{3}\int\!\! \dbar^3p f(k_0) \int\! \dbar^3k\bigg(\frac{2 k_0^2 \delta(k-k_0^T)}{|{{\partial \Re\Pi_T}/{\partial k}}|_{k_0^{T}}} \nonumber\\
&+&\frac{k^2 \delta(k-k_0^L)}{|{{\partial \Re\Pi_L}/{\partial k}}|_{k_0^{L}}}\bigg)|\langle\phi|r|o\rangle|^2.
\label{eq:decaypole}
\end{eqnarray}
The contribution of transverse gluon to the decay width of the bound states is dominant over the longitudinal one. This can be observed from the frequency behavior of $\Im\Sigma_{11}$ shown in FIG.~\ref{spec1}. Below we discuss the results obtained for the decay width and $R_{AA}$.    

\section{Results for classical dynamics~\label{sec:results}}
In this section, we discuss the results for $\Upsilon$(1S) and $\Upsilon$(2S)
states using the decay width obtained in the previous section. We mainly focus
on the relative contributions of pole and LD within the
perturbative limit and show the overall effect on $R_{AA}$. 

At any given time $t$, the survival probability of a given singlet state can be obtained by using the rate equation
\begin{equation}
\frac{d N(t)}{d t}=-N(t) \sum_i \Gamma_{i}(t),
\label{eq:rate}
\end{equation} 
where $N(t)$ is the number of bound states at time $t$ and summation is over
the two contributions arising from LD and pole. The total number of states
produced after some time $t_f$ may be obtained from Eq.~\ref{eq:rate} and is
given as \begin{equation}
N=N_0 e^{-\int_{t_0}^{t_f}\Gamma(t)dt}.
\label{eq:number}
\end{equation}
Here $t_0$ is the initial time, $N_0$ is number of bound states at time $t_0$,
and $\Gamma(t)=\sum_{i}\Gamma_{i}(t)$. 

\subsection{Medium model}
To calculate $\Gamma$ as a function of time we need a model for the background evolution of the thermal medium. In this paper we use a simple model and take the medium by a Bjorken expanding medium~\cite{Bjorken:1982qr} with a temperature 
\begin{equation}
T(t)=T(t_0) \bigg(\frac{t_0}{t}\bigg)^{\frac{1}{3}}.
\end{equation} 
We are interested in quarkonia with zero rapidity and hence have replaced the
proper time with the local time. $T(t_0)$ is the temperature at a reference
proper time $t_0$. We take $t_0=0.6$ fm which is a little later than the start
of hydrodynamics for LHC energies~\cite{Chang:2015hqa}. This is starting time for the
quarkonium evolution and is comparable to the formation time for quarkonia.
Similar numbers were taken in Ref.~\cite{brambilla20181}.

The temperature at time $t_0$ depends on the impact parameter or equivalently the centrality. We use centrality bins analogous to the one used in ALICE~\cite{ALICE:2015juo}. For this purpose, we use the Glauber model (we do not use the Monte-Carlo Glauber model here though we see that the difference between our centrality bins and the bins obtained from the Monte-Carlo Glauber model ~\cite{Miller:2007ri} is small) to relate the impact parameter to the number of participants ($N_{part}$) and the number of binary collisions($N_{bin}$). Following  ALICE, bins in the observed $dN_{ch}/d\eta$  are related to bins in $N_{part}$ using the relation
\begin{equation}
\frac{dN_{ch}}{d \eta}=\lambda(fN_{bin}+(1-f)N_{part}),
\label{eq:dnch}
\end{equation}
where $\lambda=2.75$ and $f=0.212$~\cite{ALICE:2015juo}. With these values of the parameters $\lambda$ and $f$, we quantitatively agree with the centrality dependence of  $dN_{ch}/d\eta$ as a function of $N_{part}$ given in FIG.\ref{2sraa} of Ref.\cite{ALICE:2013hur}. For each centrality bin, we have mentioned the mean value of $N_{part}$ and impact parameter $b$ in TABLE\ref{bins}. 

\begin{table}
\begin{tabular}{|c|c|c|c|c|c|c|}
\hline
Centrality(\%) & Npart & b (fm)& \multicolumn{2}{c|}{$T_0$ (MeV)}\\ 
\hline
0-2.5 & 393 & $2.5$ & 446 & 478 \\ 
2.5-5 & 363 & $3.5$ & 443 & 475 \\ 
5-7.5 & 334 & $5.0$ & 439 & 470  \\
7.5-10 & 307 & $7.0$ & 434 & 465  \\
10-20 & 248 & $8.7$ & 420 & 450  \\
20-30 & 173 & $10.0$ & 408 & 437  \\
30-40 & 116 & $11.2$ & 371 & 398  \\
40-50 & 74 & $12.2$ & 317 & 339  \\
50-60 & 44 & $13.2$ & 267 & 286 \\
60-70 & 23 & $14.0$ &207 & 222    \\
\hline
\end{tabular}
\caption{Mean value of $N_{part}$, impact parameter ($b$) and initial
temperature ($T_0$) for various centrality bins. $T_0$ is temperature at time
$t_0=0.6$ fm. The left column in $T_0$ is for $f=1.462$ and the
right one is for $f=1.782$ (see text below Eq.~\ref{eq:itemp}).}
\label{bins}
\end{table} 

With $dN_{ch}/d\eta$ in hand for each centrality bin, the initial temperature at $t_0$ is obtained by using 
following prescription~\cite{Srivastava:2016hwr}.  The value of $d\mathcal{N}/dy$  is related to the experimentally measured charged particle multiplicity by the relation,
\begin{equation}
	\frac{d\mathcal{N}}{dy}=\frac{3 J}{2} \frac{dN_{ch}}{d\eta},
	\label{eq:dndy}
\end{equation}
where $J=1.12$ is the jacobian for $y$ to $\eta$ transformation~\cite{ALICE:2016igk}. $\mathcal{N}$ is the multiplicity of the particles produced at the end, $y$ is pseudo-rapidity. The factor  $3/2$ in Eq.~\ref{eq:dndy} incorporates the contribution from neutral particles. In order to be consistent within our model, for each centrality bin, we take $dN_{ch}/d\eta$ obtained by using Eq.~\ref{eq:dnch} with the parameters mentioned above. 
 
In the nuclear collision experiments, the number density of partons or entropy is decided by the rapidity distribution of the produced particles. Assuming that initially the system is at chemical equilibrium, and assuming that the entropy does not change during the evolution, we can write  the initial parton density  as
\begin{equation}
n_0=\frac{1}{A_{\perp}t_0} \frac{d\mathcal{N}}{dy},
\label{eq:no}
\end{equation} 
where  $A_\perp$ is transverse size of the system. We estimate it from the Glauber model for each centrality via~\cite{Eyyubova:2021ngi}
\begin{equation}
A_{\perp}=4 \pi \sqrt{\langle x^2\rangle \langle y^2\rangle}, 
\label{eq:aperp}
\end{equation}
where $\langle x^2\rangle, \langle y^2\rangle $ is size along the transverse directions.  A rough estimate of $T_0$ can be made from the initial number density assuming a non-interacting QGP. Then, $n_0=(\beta_1+2 \beta_2)T_0^3$ with  $\beta_1=8\pi^2/15$ and $\beta_2=7\pi^2 N_f/40$ represents equilibrium density for gluon and quark. Plugging it back in Eq.\ref{eq:no} along with Eq.\ref{eq:dndy}, we obtain the initial temperature at time $t_0$ to be
\begin{equation}
T_0=\bigg(\frac{3 J}{2A_{\perp}t_0}\frac{dN_{ch}}{d\eta}\frac{1}{\beta_1+2 \beta_2}\bigg)^{\frac{1}{3}}.
\label{eq:itemp}
\end{equation}

For $\sqrt{s}=2.76$ TeV, Eq.\ref{eq:itemp} gives $T_0\sim250$ MeV for the most
central bin ($0-2.5$\%).  This is significantly lower than estimates of the
temperatures obtained at $t_0 \approx 0.6$ fm in hydrodynamic
simulations~\cite{Chang:2015hqa}.  We note that Eq.\ref{eq:no} ignores various
effects such as particalization, viscous effects, interaction in the QGP, and
expansion in the transverse direction. Moreover, losses in the work done by the
system during the expansion are also not taken into account. These effects may
lead to entropy generation and energy loss restricting the applicability of
Bjorken expansion. Thus the initial parton density will be somewhat larger than
estimated by Eq.~\ref{eq:no}. In order to take these losses into account we
redefine the initial parton density by multiplying Eq.~\ref{eq:no} with a fudge
factor ($f$) i.e., $n_0=f n_0$. 
\begin{figure}[h]
	\centering 
	\includegraphics[width=.43\textwidth,origin=c,angle=0]{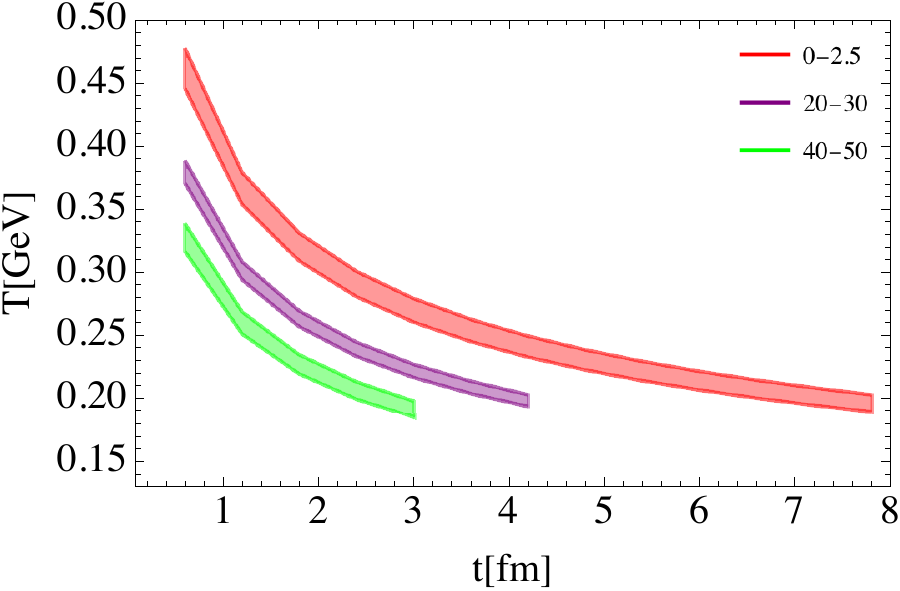}
	\caption{\label{tempev} Temperature as a function of time. Red band is for
		the most central bin and, central and green one are for 20-30\% and 40-50\%  centrality, respectively. Here
		the initial time $t_0=0.6$ fm for all centrality bins. The band here covers upper and lower temperature  correspond to two different values of the fudge factor. }
\end{figure}
This fudge factor is adjusted in such a
way that we obtain the initial temperature in the most central bin to range from
$T_0\approx450$ MeV  ($f=1.46$) to $T_0\approx480$ MeV ($f=1.78$) at $t_0=0.6$
fm.  For our final results of $R_{AA}$ we provide the band corresponding to these
two values of the initial temperatures. While this is a highly simplified model
for the background medium, we hope that this band of variation captures
important features of its hydrodynamic evolution.

In FIG.\ref{tempev}, we show the variation of  temperature as a function of
time starting with $t_0=0.6$ fm till the temperature reaches the final value of
190 MeV.

Finally, for completeness, we also consider the final state feed-down effect,
we follow the prescription given in Ref.~\cite{Islam:2020bnp}. Therefore, for a
given bound states, $R_{AA}$  is given by
\begin{equation}
	R_{AA}=\frac{N+\alpha N_h}{N_0+\alpha N_{0h}},
	\label{eq:raa}
\end{equation}
where $N_h$ is number of higher states at the end of evolution and $\alpha$ is
feed-down parameter from higher state ($N_h$) to the state being considered.
Feed-down matrix is given in Eq.(5.2) of Ref.\cite{Islam:2020bnp}.  $N_0$ is
number of states at the initial time $t_0$. In the results presented here, for
feed-down , we only consider higher states with contributions more than 10\%.
Thus for $\Upsilon$(1S), we take the contribution from $\Upsilon$(2S),
$\chi_{b0}$(1P),$\chi_{b1}$(1P) and $\chi_{b1}$(2P) states. We follow the same
for $\Upsilon$(2S) state as well.  For more details on feed down see
Ref.\cite{Islam:2020bnp}. Below we discuss the results in somewhat more detail.

\subsection{Results}
\begin{figure}[h]
	\centering 
	\includegraphics[width=.43\textwidth,origin=c,angle=0]{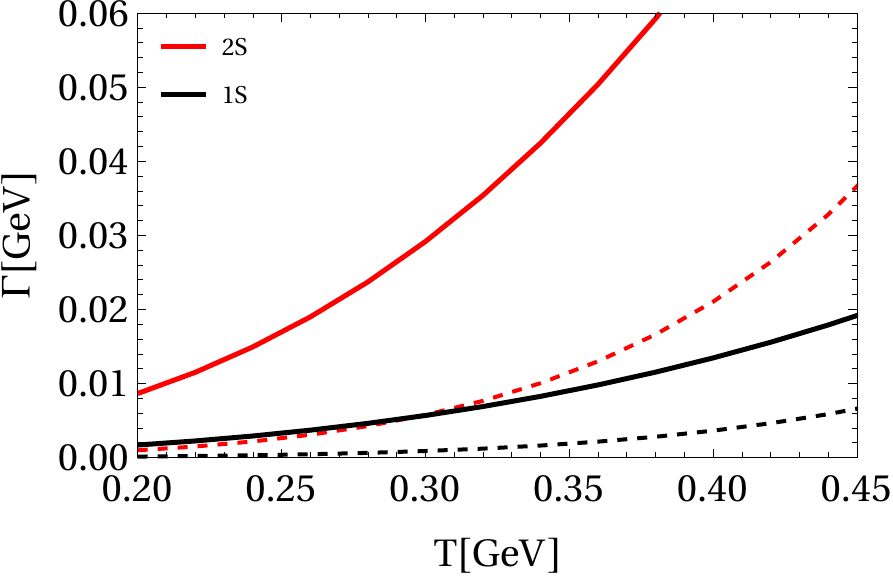}
	\caption{Decay width of 1S (black) and 2S (red) from imaginary potential (solid line) and longitudinal LD (dashed line) given in Eq.\ref{eq:LD}. }
	\label{fig:kcomp}
\end{figure}

In FIG.~\ref{fig:kcomp}, we  compare the decay width (of $\Upsilon$(1S) and
$\Upsilon$(2S) states) from Eq.~\ref{eq:LD} with the (imaginary)
potential~\cite{laine20072} formalism which is extensively used in the
literature. In the latter case the decay width is obtained from
\begin{equation}
\Gamma=2\langle \phi |\Im V_s(r,T)|\phi\rangle, 
\label{eq:dpot}
\end{equation}
where $\Im V_s$ is the imaginary part of the singlet potential.  In order to be
consistent with the $r^2$ expansion in Eq.~\ref{eq:LpNRQCD}, we also make this
approximation in the imaginary part of the potential.  In this limit, the
momentum integration in the potential becomes divergent, limiting its
applicability in the small momentum ranges. We therefore put an upper cut
off~$m_D$ to get finite results. The corresponding decay width for
$\Upsilon$(1S) and $\Upsilon$(2S) states are shown by the solid line in
FIG.~\ref{fig:kcomp}. In the same figure, we show the decay width obtained from
Eq.~\ref{eq:LD} with $E_b=0.6$ GeV (for 1S) and $0.2$ GeV for 2S. As may be
noticed, $\Im V_s(r,T)$ gives a larger decay width  which can be understood as
follows. 

\begin{figure}[h]
	\centering 
	\includegraphics[width=.43\textwidth,origin=c,angle=0]{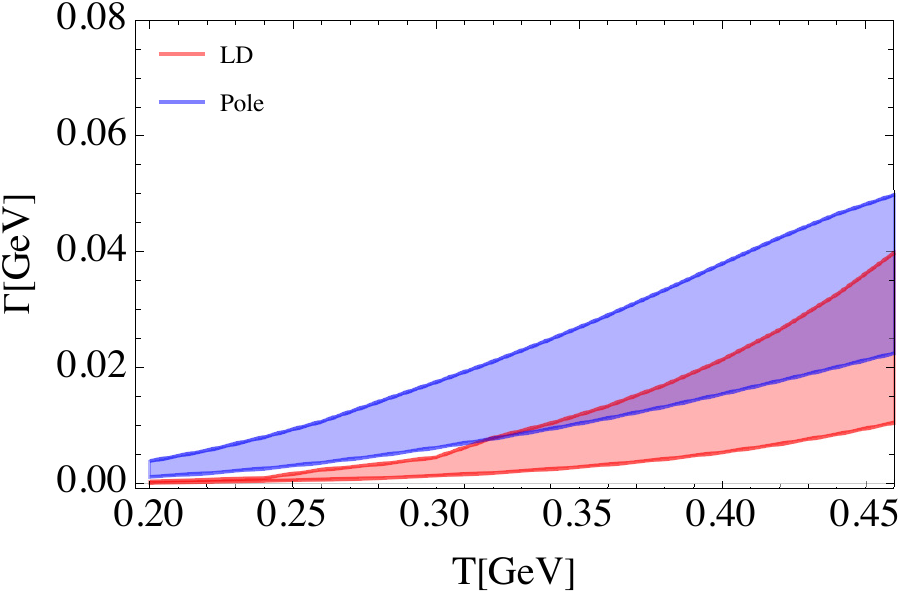}
	\caption{Decay width of the $\Upsilon(1{\rm{S}})$  state as a function of medium temperature. The band corresponds to the free wave function (upper curve) (Eq.~\ref{eq:wavef}) and Coloumbic repulsive wavefunction (lower curve) (Eq.~\ref{eq:wavefn}). Red color band is LD and blue color band is gluo-dissociation.  }
	\label{fig:comp}
\end{figure} 

Eq.~\ref{eq:dpot} assumes the binding energy of
the singlet state, and the octet state energy ($q^2/M$) of the species is
negligible compared to the temperature.  This can be seen by inserting a
complete set of octet states in Eq.~\ref{eq:LD}. The energy delta function
gives $k_0=q^2/M-E_b$. Taking the limit $k_0\to 0$, Eq.\ref{eq:LD} gives the
result obtained from the imaginary potential expanded in $r$ to order $2$. We
emphasize here that for realistic values, this approximation over-predicts the
decay rate. We have checked that if we do not make an expansion in $r$ in the
imaginary potential, the decay width turns out to be even larger. 

\begin{figure}[h]
	\centering 
	\includegraphics[width=.43\textwidth,origin=c,angle=0]{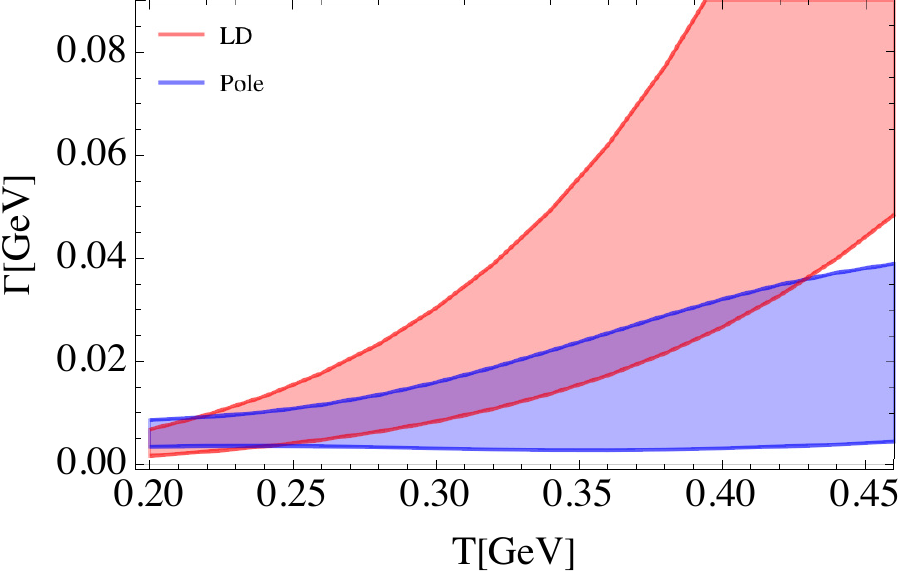}
	\caption{Decay width of the $\Upsilon(2{\rm{S}})$   state as a function of medium temperature. The  color conventions are the same as in Fig.~\ref{fig:comp}}
	\label{fig:comp2s}
\end{figure}

In FIG.~\ref{fig:comp}, we plot the scattering (LD) and gluo dissociation
(pole) contributions to $\Upsilon$(1S) state. For $\Upsilon$(2S) we plot the
same quantity in FIG.~\ref{fig:comp2s}. The bands correspond to no screening
(Eq.~\ref{eq:wavefn}) and complete screening (Eq.~\ref{eq:wavef}) scenarios.
Here we take $E_b=0.6$ GeV (for 1S) and $E_b=0.2$ GeV (for 2S) which
lie in the middle (see FIG.~\ref{be}) of the binding energies within the
temperature range of interest.

It may be observed (from FIG.~\ref{be}) that the hierarchies $E_b\gg T$ or
$E_b\ll T$ are not very well satisfied for both $1S$ and $2S$ states. One would
therefore expect significant contribution from the finite frequency region of
FIG.~\ref{fig:spectral}.   For $\Upsilon$(1S), this  may be observed in
FIG.~\ref{fig:comp}. As anticipated, with this value of $E_b$, pole (blue band)
and LD (red band) give somewhat similar contribution for a wide range of
temperature. 
\begin{figure}[h]
	\centering 
	\includegraphics[width=.41\textwidth,origin=c,angle=0]{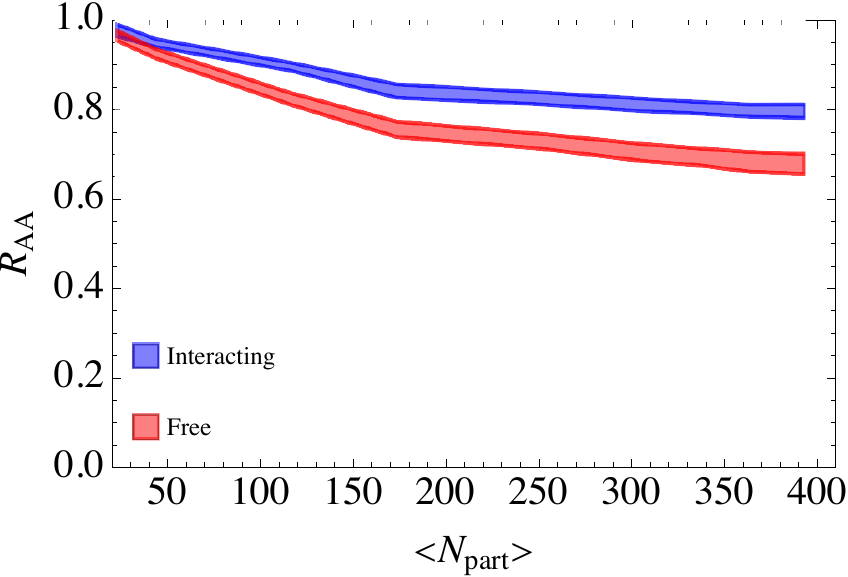}
	\caption{\label{1sraa} $R_{AA}$ for $\Upsilon(1S)$ state as a function of $\langle N_{part}\rangle $ for each centrality bin.  For blue color we take octet potential as vacuum potential and for red color we switch off octet potential assuming complete screening.}
\end{figure}
Moreover, LD contribution dominates at high temperature. Similarly, for
$\Upsilon$(2S) state, pole  contribution is larger at temperature $\sim 200$
MeV and at high temperature $\sim 400$ MeV, LD contribution is significantly
larger than the pole contribution. This suggest that for $\Upsilon$(2S)
dissociation, LD gives dominant contribution. 

\begin{figure}[h]
\centering 
\includegraphics[width=.41\textwidth,origin=c,angle=0]{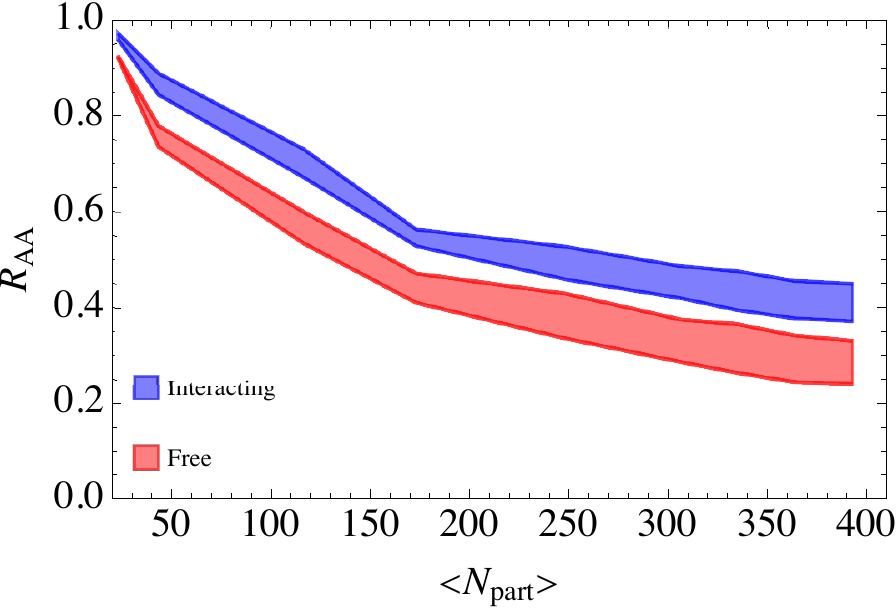}
\caption{\label{2sraa}$R_{AA}$ for $\Upsilon(2S)$ state as a function of $<N_{part}>$ for each centrality bin.  Color representation is same as FIG.\ref{1sraa}.}
\end{figure}

In FIG.~\ref{1sraa}, we combine both gluon absorption and scattering processes
and plot $R_{AA}$ for $\Upsilon(1S) $ state as a function of $\langle
N_{part}\rangle$ for each centrality bin. 

For a given centrality, we first evaluate  singlet state wave functions at each
temperature by solving the Schrodinger equation given in Eq.\ref{eq:singlet}.
With temperature dependent wave functions at hand, we estimate matrix element
i.e.,  $|\langle \phi|r|o\rangle|^2$ (see Eq.\ref{eq:LD}) and obtain
corresponding decay width for all processes.  Here we use temperature dependent
binding energy shown in FIG.~\ref{be}. Finally, we use Eq.~\ref{eq:raa} to
estimate  $R_{AA}$ for a given state. As mentioned earlier, we consider two
cases ; (a) when there is no screening in the final state octet interaction
(show in blue color)and (b) when octet states are completely screened (shown in
red color). Here, upper and lower edge of the band corresponds to the lower and
higher initial temperature given in TABLE~\ref{bins}. 

Let us note that in the realistic situations,  octet potential is neither
completely screened nor has form of pure vacuum potential. Therefore, we expect
the true value of suppression to lie somewhere between the blue and the red bands. It
is evident that the suppression is under-predicted
for these states.  This may largely be due to the fact that the spectral
functions are obtained at the leading order (LO)  accuracy. Further
improvements of our results require non-perturbative estimate of these spectral
functions on lattice which so far have been computed only in the static limit.

\section{Conclusion~\label{sec:conclusion}}
In this work, we perform a comprehensive analysis to quantify the relative
contributions arising from LD and gluo-dissociation to the $\Upsilon$(1S) and
$\Upsilon$(2S) states within leading order perturbation theory for the medium.
We assume that the real part of the singlet potential is screened in a thermal
medium, and we take lattice motivated real parts of the $Q\bar{Q}$ potential
and estimate the binding energy and wave function as a function of $T$ for
$\Upsilon$ and $\chi$ states. By comparing the binding energy (shown in
FIG.~\ref{be}) and temperature it can be seen that for the 1S and 2S states
both $E_b$ and $T$ are comparable to each other (although for the 2S state for
$T>300$MeV one could take $T\gg E_b$). We, therefore, expect that neither $E_b\gg
T$ nor $E_b\ll T$ hierarchy is strictly satisfied in the interesting range of
temperatures for the QGP, at least for 1S and 2S. However, for other higher
excited states, $E_b\ll T$ seems to be satisfied quite well.  For $\Upsilon(3S)$
this can be seen in FIG.~\ref{be}.

The consequence for the 1S state (and for the 2S in the later stages of the
evolution) is that LD can not adequately describe the decay of the state. It is
well known that in the static limit i.e., $k_0\rightarrow 0$, decay dominantly
happens via longitudinal LD which can be captured by an imaginary potential
between $Q\bar{Q}$. However, this limit is valid if the binding energy of the
quarkonium species is negligible compared to the medium temperature. As
discussed in the text below FIG.~\ref{fig:kcomp}, this also requires the kinetic
energy of the octet states to be small. For $E_b\sim T$, the full finite
frequency region of the spectral function for the chromo-electric field
correlator  becomes important.

There is a caveat to the above conclusion. The perturbative value of the
$k_0\rightarrow 0$ value of the chromo-electric spectral function is a factor
of $5-10$ times smaller than the non-perturbative value  calculated on the
lattice.  If this enhancement persists till $k_0$ values of a few 100 MeV (see
FIG.~\ref{fig:spectral}) then our conclusion about the relative contribution of
the LD and the other contributions will  be still be true. However, if the
spectral function rapidly drops down towards the perturbative estimate (at
$k_0\sim$ GeV we expect the leading order perturbative result to be more reliable) then the
LD contribution dominates for all the states and the static results can be used
to capture the physics of interest. This provides a motivation to study the
spectral function at finite frequency non-perturbatively, although it is a
difficult problem.

The hierarchy between $E_b$ and $T$ ($T$ is related to the inverse of the
environment time scale) plays an important role while performing a quantum
calculations of quarkonia dynamics within the open quantum system framework.
It has been shown~\cite{akamatsu20151,Akamatsu:2020ypb,brambilla20171} that if
$E_b \ll T$ quarkonium system is in the quantum Brownian motion regime and the
evolution is local in time. In this case, one can obtain a Lindblad equation
for the density matrix evolution for the $Q\bar{Q}$ system.  On the other hand
if $E_b\gtrsim T$ then this proof fails. We show here that this is the case for
the $\Upsilon(1\rm{S})$ state throughout the evolution if one considers the
state to be an eigenstate of the instantaneous (real) potential.  $E_b$ is also
$\lesssim T$ for the excited states in the early part of the evolution.
However, If substantial  regeneration of the excited states happens at the
later time (low temperature) then  the hierarchy $E_b\gtrsim T$  might be
satisfied. In this regime if a quantum Brownian description of the dynamics is
attempted, one might need dynamics that are correlated in time. Although a
quantum optical description might become applicable in this
case~\cite{yao20191,Yao:2020eqy}. We do not discuss quantum evolution in this
work and leave this for the future. 

\section{\label{sec:acknowledgements}Acknowledgements}
We acknowledge support of the Department of Atomic Energy, Government of
India, under Project Identification No. RTI 4002. 
We also thank Saumen Datta
and Michael Strickland for valuable discussions.

\section{\label{sec:references}References}
\bibliography{Hierarchies}

\providecommand{\noopsort}[1]{}\providecommand{\singleletter}[1]{#1}%
\begin{thebibliography}{88}%
\makeatletter
\providecommand \@ifxundefined [1]{%
 \@ifx{#1\undefined}
}%
\providecommand \@ifnum [1]{%
 \ifnum #1\expandafter \@firstoftwo
 \else \expandafter \@secondoftwo
 \fi
}%
\providecommand \@ifx [1]{%
 \ifx #1\expandafter \@firstoftwo
 \else \expandafter \@secondoftwo
 \fi
}%
\providecommand \natexlab [1]{#1}%
\providecommand \enquote  [1]{``#1''}%
\providecommand \bibnamefont  [1]{#1}%
\providecommand \bibfnamefont [1]{#1}%
\providecommand \citenamefont [1]{#1}%
\providecommand \href@noop [0]{\@secondoftwo}%
\providecommand \href [0]{\begingroup \@sanitize@url \@href}%
\providecommand \@href[1]{\@@startlink{#1}\@@href}%
\providecommand \@@href[1]{\endgroup#1\@@endlink}%
\providecommand \@sanitize@url [0]{\catcode `\\12\catcode `\$12\catcode
  `\&12\catcode `\#12\catcode `\^12\catcode `\_12\catcode `\%12\relax}%
\providecommand \@@startlink[1]{}%
\providecommand \@@endlink[0]{}%
\providecommand \url  [0]{\begingroup\@sanitize@url \@url }%
\providecommand \@url [1]{\endgroup\@href {#1}{\urlprefix }}%
\providecommand \urlprefix  [0]{URL }%
\providecommand \Eprint [0]{\href }%
\providecommand \doibase [0]{http://dx.doi.org/}%
\providecommand \selectlanguage [0]{\@gobble}%
\providecommand \bibinfo  [0]{\@secondoftwo}%
\providecommand \bibfield  [0]{\@secondoftwo}%
\providecommand \translation [1]{[#1]}%
\providecommand \BibitemOpen [0]{}%
\providecommand \bibitemStop [0]{}%
\providecommand \bibitemNoStop [0]{.\EOS\space}%
\providecommand \EOS [0]{\spacefactor3000\relax}%
\providecommand \BibitemShut  [1]{\csname bibitem#1\endcsname}%
\let\auto@bib@innerbib\@empty
\bibitem [{\citenamefont {{N. Brambilla, A. Pineda, J. Soto and A.
  Vairo}}(2000)}]{brambilla20001}%
  \BibitemOpen
  \bibfield  {author} {\bibinfo {author} {\bibnamefont {{N. Brambilla, A.
  Pineda, J. Soto and A. Vairo}}},\ }\href {\doibase
  https://doi.org/10.1016/S0550-3213(99)00693-8} {\bibfield  {journal}
  {\bibinfo  {journal} {Nuclear Physics B}\ }\textbf {\bibinfo {volume}
  {566}},\ \bibinfo {pages} {275 } (\bibinfo {year} {2000})}\BibitemShut
  {NoStop}%
\bibitem [{\citenamefont {{A. Pineda and J. Soto}}(1998)}]{pineda19981}%
  \BibitemOpen
  \bibfield  {author} {\bibinfo {author} {\bibnamefont {{A. Pineda and J.
  Soto}}},\ }\href {\doibase https://doi.org/10.1016/S0920-5632(97)01102-X}
  {\bibfield  {journal} {\bibinfo  {journal} {Nuclear Physics B - Proceedings
  Supplements}\ }\textbf {\bibinfo {volume} {64}},\ \bibinfo {pages} {428 }
  (\bibinfo {year} {1998})},\ \bibinfo {note} {proceedings of the QCD 97
  Euroconference 25th Anniversary of QCD}\BibitemShut {NoStop}%
\bibitem [{\citenamefont {{T. Matsui and H. Satz}}(1986)}]{matsui19861}%
  \BibitemOpen
  \bibfield  {author} {\bibinfo {author} {\bibnamefont {{T. Matsui and H.
  Satz}}},\ }\href {\doibase 10.1016/0370-2693(86)91404-8} {\bibfield
  {journal} {\bibinfo  {journal} {Phys. Lett.}\ }\textbf {\bibinfo {volume}
  {B178}},\ \bibinfo {pages} {416} (\bibinfo {year} {1986})}\BibitemShut
  {NoStop}%
\bibitem [{\citenamefont {Laine}\ \emph {et~al.}(2007)\citenamefont {Laine},
  \citenamefont {Philipsen}, \citenamefont {Tassler},\ and\ \citenamefont
  {Romatschke}}]{laine20071}%
  \BibitemOpen
  \bibfield  {author} {\bibinfo {author} {\bibfnamefont {M.}~\bibnamefont
  {Laine}}, \bibinfo {author} {\bibfnamefont {O.}~\bibnamefont {Philipsen}},
  \bibinfo {author} {\bibfnamefont {M.}~\bibnamefont {Tassler}}, \ and\
  \bibinfo {author} {\bibfnamefont {P.}~\bibnamefont {Romatschke}},\ }\href
  {\doibase 10.1088/1126-6708/2007/03/054} {\bibfield  {journal} {\bibinfo
  {journal} {Journal of High Energy Physics}\ }\textbf {\bibinfo {volume}
  {2007}},\ \bibinfo {pages} {054} (\bibinfo {year} {2007})}\BibitemShut
  {NoStop}%
\bibitem [{\citenamefont {{G. Bhanot and M. E. Peskin }}(1979)}]{bhanot19791}%
  \BibitemOpen
  \bibfield  {author} {\bibinfo {author} {\bibnamefont {{G. Bhanot and M. E.
  Peskin }}},\ }\href {\doibase 10.1016/0550-3213(79)90200-1} {\bibfield
  {journal} {\bibinfo  {journal} {Nucl. Phys.}\ }\textbf {\bibinfo {volume}
  {B156}},\ \bibinfo {pages} {391} (\bibinfo {year} {1979})}\BibitemShut
  {NoStop}%
\bibitem [{\citenamefont {{N. Brambilla, J. Ghiglieri, A. Vairo and P.
  Petreczky}}(2008)}]{brambilla20081}%
  \BibitemOpen
  \bibfield  {author} {\bibinfo {author} {\bibnamefont {{N. Brambilla, J.
  Ghiglieri, A. Vairo and P. Petreczky}}},\ }\href {\doibase
  10.1103/PhysRevD.78.014017} {\bibfield  {journal} {\bibinfo  {journal} {Phys.
  Rev. D}\ }\textbf {\bibinfo {volume} {78}},\ \bibinfo {pages} {014017}
  (\bibinfo {year} {2008})}\BibitemShut {NoStop}%
\bibitem [{\citenamefont {Peskin}(1979)}]{peskin19791}%
  \BibitemOpen
  \bibfield  {author} {\bibinfo {author} {\bibfnamefont {M.~E.}\ \bibnamefont
  {Peskin}},\ }\href {\doibase https://doi.org/10.1016/0550-3213(79)90199-8}
  {\bibfield  {journal} {\bibinfo  {journal} {Nuclear Physics B}\ }\textbf
  {\bibinfo {volume} {156}},\ \bibinfo {pages} {365 } (\bibinfo {year}
  {1979})}\BibitemShut {NoStop}%
\bibitem [{\citenamefont {Braaten}\ and\ \citenamefont
  {Pisarski}(1990)}]{Braaten:1990it}%
  \BibitemOpen
  \bibfield  {author} {\bibinfo {author} {\bibfnamefont {E.}~\bibnamefont
  {Braaten}}\ and\ \bibinfo {author} {\bibfnamefont {R.~D.}\ \bibnamefont
  {Pisarski}},\ }\href {\doibase 10.1103/PhysRevD.42.2156} {\bibfield
  {journal} {\bibinfo  {journal} {Phys. Rev. D}\ }\textbf {\bibinfo {volume}
  {42}},\ \bibinfo {pages} {2156} (\bibinfo {year} {1990})}\BibitemShut
  {NoStop}%
\bibitem [{\citenamefont {Kaczmarek}\ \emph {et~al.}(2004)\citenamefont
  {Kaczmarek}, \citenamefont {Ejiri}, \citenamefont {Karsch}, \citenamefont
  {Laermann},\ and\ \citenamefont {Zantow}}]{kaczmarek20031}%
  \BibitemOpen
  \bibfield  {author} {\bibinfo {author} {\bibfnamefont {O.}~\bibnamefont
  {Kaczmarek}}, \bibinfo {author} {\bibfnamefont {S.}~\bibnamefont {Ejiri}},
  \bibinfo {author} {\bibfnamefont {F.}~\bibnamefont {Karsch}}, \bibinfo
  {author} {\bibfnamefont {E.}~\bibnamefont {Laermann}}, \ and\ \bibinfo
  {author} {\bibfnamefont {F.}~\bibnamefont {Zantow}},\ }\bibfield  {booktitle}
  {\emph {\bibinfo {booktitle} {{Finite density QCD. Proceedings, International
  Workshop, Nara, Japan, July 10-12, 2003}}},\ }\href {\doibase
  10.1143/PTPS.153.287} {\bibfield  {journal} {\bibinfo  {journal} {Prog.
  Theor. Phys. Suppl.}\ }\textbf {\bibinfo {volume} {153}},\ \bibinfo {pages}
  {287} (\bibinfo {year} {2004})},\ \Eprint
  {http://arxiv.org/abs/hep-lat/0312015} {arXiv:hep-lat/0312015 [hep-lat]}
  \BibitemShut {NoStop}%
\bibitem [{\citenamefont {Brambilla}\ \emph {et~al.}(2021)\citenamefont
  {Brambilla}, \citenamefont {Escobedo}, \citenamefont {Strickland},
  \citenamefont {Vairo}, \citenamefont {Vander~Griend},\ and\ \citenamefont
  {Weber}}]{Brambilla:2020qwo}%
  \BibitemOpen
  \bibfield  {author} {\bibinfo {author} {\bibfnamefont {N.}~\bibnamefont
  {Brambilla}}, \bibinfo {author} {\bibfnamefont {M.~A.}\ \bibnamefont
  {Escobedo}}, \bibinfo {author} {\bibfnamefont {M.}~\bibnamefont
  {Strickland}}, \bibinfo {author} {\bibfnamefont {A.}~\bibnamefont {Vairo}},
  \bibinfo {author} {\bibfnamefont {P.}~\bibnamefont {Vander~Griend}}, \ and\
  \bibinfo {author} {\bibfnamefont {J.~H.}\ \bibnamefont {Weber}},\ }\href
  {\doibase 10.1007/JHEP05(2021)136} {\bibfield  {journal} {\bibinfo  {journal}
  {JHEP}\ }\textbf {\bibinfo {volume} {05}},\ \bibinfo {pages} {136} (\bibinfo
  {year} {2021})},\ \Eprint {http://arxiv.org/abs/2012.01240} {arXiv:2012.01240
  [hep-ph]} \BibitemShut {NoStop}%
\bibitem [{\citenamefont {{S. Caron-Huot and G. D. Moore}}(2008)}]{moore20081}%
  \BibitemOpen
  \bibfield  {author} {\bibinfo {author} {\bibnamefont {{S. Caron-Huot and G.
  D. Moore}}},\ }\href {\doibase 10.1103/PhysRevLett.100.052301} {\bibfield
  {journal} {\bibinfo  {journal} {Phys. Rev. Lett.}\ }\textbf {\bibinfo
  {volume} {100}},\ \bibinfo {pages} {052301} (\bibinfo {year}
  {2008})}\BibitemShut {NoStop}%
\bibitem [{\citenamefont {Andronic}\ \emph {et~al.}(2016)\citenamefont
  {Andronic} \emph {et~al.}}]{andronic20151}%
  \BibitemOpen
  \bibfield  {author} {\bibinfo {author} {\bibfnamefont {A.}~\bibnamefont
  {Andronic}} \emph {et~al.},\ }\href {\doibase 10.1140/epjc/s10052-015-3819-5}
  {\bibfield  {journal} {\bibinfo  {journal} {Eur. Phys. J.}\ }\textbf
  {\bibinfo {volume} {C76}},\ \bibinfo {pages} {107} (\bibinfo {year}
  {2016})},\ \Eprint {http://arxiv.org/abs/1506.03981} {arXiv:1506.03981
  [nucl-ex]} \BibitemShut {NoStop}%
\bibitem [{\citenamefont {{ L. Grandchamp, R. Rapp and G. E.
  Brown}}(2004)}]{grandchamp20041}%
  \BibitemOpen
  \bibfield  {author} {\bibinfo {author} {\bibnamefont {{ L. Grandchamp, R.
  Rapp and G. E. Brown}}},\ }\href {\doibase 10.1103/PhysRevLett.92.212301}
  {\bibfield  {journal} {\bibinfo  {journal} {Phys. Rev. Lett.}\ }\textbf
  {\bibinfo {volume} {92}},\ \bibinfo {pages} {212301} (\bibinfo {year}
  {2004})},\ \Eprint {http://arxiv.org/abs/hep-ph/0306077}
  {arXiv:hep-ph/0306077 [hep-ph]} \BibitemShut {NoStop}%
\bibitem [{\citenamefont {{ R. Rapp and H. van Hees}}(2010)}]{rapp20091}%
  \BibitemOpen
  \bibfield  {author} {\bibinfo {author} {\bibnamefont {{ R. Rapp and H. van
  Hees}}},\ }in\ \href {\doibase 10.1142/9789814293297_0003} {\emph {\bibinfo
  {booktitle} {{Quark-gluon plasma 4}}}}\ (\bibinfo {year} {2010})\ pp.\
  \bibinfo {pages} {111--206},\ \Eprint {http://arxiv.org/abs/0903.1096}
  {arXiv:0903.1096 [hep-ph]} \BibitemShut {NoStop}%
\bibitem [{\citenamefont {{X. Zhao and R. Rapp}}(2011)}]{zhao20111}%
  \BibitemOpen
  \bibfield  {author} {\bibinfo {author} {\bibnamefont {{X. Zhao and R.
  Rapp}}},\ }\href {\doibase 10.1016/j.nuclphysa.2011.05.001} {\bibfield
  {journal} {\bibinfo  {journal} {Nucl. Phys.}\ }\textbf {\bibinfo {volume}
  {A859}},\ \bibinfo {pages} {114} (\bibinfo {year} {2011})},\ \Eprint
  {http://arxiv.org/abs/1102.2194} {arXiv:1102.2194 [hep-ph]} \BibitemShut
  {NoStop}%
\bibitem [{\citenamefont {Emerick}\ \emph {et~al.}(2012)\citenamefont
  {Emerick}, \citenamefont {Zhao},\ and\ \citenamefont {Rapp}}]{emerick20111}%
  \BibitemOpen
  \bibfield  {author} {\bibinfo {author} {\bibfnamefont {A.}~\bibnamefont
  {Emerick}}, \bibinfo {author} {\bibfnamefont {X.}~\bibnamefont {Zhao}}, \
  and\ \bibinfo {author} {\bibfnamefont {R.}~\bibnamefont {Rapp}},\ }\href
  {\doibase 10.1140/epja/i2012-12072-y} {\bibfield  {journal} {\bibinfo
  {journal} {Eur. Phys. J.}\ }\textbf {\bibinfo {volume} {A48}},\ \bibinfo
  {pages} {72} (\bibinfo {year} {2012})},\ \Eprint
  {http://arxiv.org/abs/1111.6537} {arXiv:1111.6537 [hep-ph]} \BibitemShut
  {NoStop}%
\bibitem [{\citenamefont {Zhao}\ \emph {et~al.}(2013)\citenamefont {Zhao},
  \citenamefont {Emerick},\ and\ \citenamefont {Rapp}}]{zhao20121}%
  \BibitemOpen
  \bibfield  {author} {\bibinfo {author} {\bibfnamefont {X.}~\bibnamefont
  {Zhao}}, \bibinfo {author} {\bibfnamefont {A.}~\bibnamefont {Emerick}}, \
  and\ \bibinfo {author} {\bibfnamefont {R.}~\bibnamefont {Rapp}},\ }\bibfield
  {booktitle} {\emph {\bibinfo {booktitle} {{Proceedings, 23rd International
  Conference on Ultrarelativistic Nucleus-Nucleus Collisions : Quark Matter
  2012 (QM 2012): Washington, DC, USA, August 13-18, 2012}}},\ }\href {\doibase
  10.1016/j.nuclphysa.2013.02.088} {\bibfield  {journal} {\bibinfo  {journal}
  {Nucl. Phys.}\ }\textbf {\bibinfo {volume} {A904-905}},\ \bibinfo {pages}
  {611c} (\bibinfo {year} {2013})},\ \Eprint {http://arxiv.org/abs/1210.6583}
  {arXiv:1210.6583 [hep-ph]} \BibitemShut {NoStop}%
\bibitem [{\citenamefont {Du}\ \emph {et~al.}(2017{\natexlab{a}})\citenamefont
  {Du}, \citenamefont {He},\ and\ \citenamefont {Rapp}}]{du20171}%
  \BibitemOpen
  \bibfield  {author} {\bibinfo {author} {\bibfnamefont {X.}~\bibnamefont
  {Du}}, \bibinfo {author} {\bibfnamefont {M.}~\bibnamefont {He}}, \ and\
  \bibinfo {author} {\bibfnamefont {R.}~\bibnamefont {Rapp}},\ }\href {\doibase
  10.1103/PhysRevC.96.054901} {\bibfield  {journal} {\bibinfo  {journal} {Phys.
  Rev.}\ }\textbf {\bibinfo {volume} {C96}},\ \bibinfo {pages} {054901}
  (\bibinfo {year} {2017}{\natexlab{a}})},\ \Eprint
  {http://arxiv.org/abs/1706.08670} {arXiv:1706.08670 [hep-ph]} \BibitemShut
  {NoStop}%
\bibitem [{\citenamefont {{ X. Du and R. Rapp}}(2019)}]{du20181}%
  \BibitemOpen
  \bibfield  {author} {\bibinfo {author} {\bibnamefont {{ X. Du and R.
  Rapp}}},\ }\href {\doibase 10.1007/JHEP03(2019)015} {\bibfield  {journal}
  {\bibinfo  {journal} {JHEP}\ }\textbf {\bibinfo {volume} {03}},\ \bibinfo
  {pages} {015} (\bibinfo {year} {2019})},\ \Eprint
  {http://arxiv.org/abs/1808.10014} {arXiv:1808.10014 [nucl-th]} \BibitemShut
  {NoStop}%
\bibitem [{\citenamefont {{F. Brezinski and G.
  Wolschin}}(2012)}]{brezinski20121}%
  \BibitemOpen
  \bibfield  {author} {\bibinfo {author} {\bibnamefont {{F. Brezinski and G.
  Wolschin}}},\ }\href {\doibase
  https://doi.org/10.1016/j.physletb.2012.01.012} {\bibfield  {journal}
  {\bibinfo  {journal} {Physics Letters B}\ }\textbf {\bibinfo {volume}
  {707}},\ \bibinfo {pages} {534 } (\bibinfo {year} {2012})}\BibitemShut
  {NoStop}%
\bibitem [{\citenamefont {{F. Nendzig and G. Wolschin}}(2013)}]{nendzig20131}%
  \BibitemOpen
  \bibfield  {author} {\bibinfo {author} {\bibnamefont {{F. Nendzig and G.
  Wolschin}}},\ }\href {\doibase 10.1103/PhysRevC.87.024911} {\bibfield
  {journal} {\bibinfo  {journal} {Phys. Rev. C}\ }\textbf {\bibinfo {volume}
  {87}},\ \bibinfo {pages} {024911} (\bibinfo {year} {2013})}\BibitemShut
  {NoStop}%
\bibitem [{\citenamefont {{J. Hong and H. Su Lee}}(2019)}]{hong20191}%
  \BibitemOpen
  \bibfield  {author} {\bibinfo {author} {\bibnamefont {{J. Hong and H. Su
  Lee}}},\ }\href@noop {} {\  (\bibinfo {year} {2019})},\ \Eprint
  {http://arxiv.org/abs/1909.07696} {arXiv:1909.07696 [nucl-th]} \BibitemShut
  {NoStop}%
\bibitem [{\citenamefont {{Michael Strickland}}(2011)}]{strickland20111}%
  \BibitemOpen
  \bibfield  {author} {\bibinfo {author} {\bibnamefont {{Michael
  Strickland}}},\ }\href {\doibase 10.1103/PhysRevLett.107.132301} {\bibfield
  {journal} {\bibinfo  {journal} {Phys. Rev. Lett.}\ }\textbf {\bibinfo
  {volume} {107}},\ \bibinfo {pages} {132301} (\bibinfo {year} {2011})},\
  \Eprint {http://arxiv.org/abs/1106.2571} {arXiv:1106.2571 [hep-ph]}
  \BibitemShut {NoStop}%
\bibitem [{\citenamefont {{M. Strickland and D.
  Bazow}}(2012)}]{strickland20112}%
  \BibitemOpen
  \bibfield  {author} {\bibinfo {author} {\bibnamefont {{M. Strickland and D.
  Bazow}}},\ }\href {\doibase 10.1016/j.nuclphysa.2012.02.003} {\bibfield
  {journal} {\bibinfo  {journal} {Nucl. Phys.}\ }\textbf {\bibinfo {volume}
  {A879}},\ \bibinfo {pages} {25} (\bibinfo {year} {2012})},\ \Eprint
  {http://arxiv.org/abs/1112.2761} {arXiv:1112.2761 [nucl-th]} \BibitemShut
  {NoStop}%
\bibitem [{\citenamefont {Margotta}\ \emph {et~al.}(2011)\citenamefont
  {Margotta}, \citenamefont {McCarty}, \citenamefont {McGahan}, \citenamefont
  {Strickland},\ and\ \citenamefont {Yager-Elorriaga}}]{margotta20111}%
  \BibitemOpen
  \bibfield  {author} {\bibinfo {author} {\bibfnamefont {M.}~\bibnamefont
  {Margotta}}, \bibinfo {author} {\bibfnamefont {K.}~\bibnamefont {McCarty}},
  \bibinfo {author} {\bibfnamefont {C.}~\bibnamefont {McGahan}}, \bibinfo
  {author} {\bibfnamefont {M.}~\bibnamefont {Strickland}}, \ and\ \bibinfo
  {author} {\bibfnamefont {D.}~\bibnamefont {Yager-Elorriaga}},\ }\href
  {\doibase 10.1103/PhysRevD.84.069902, 10.1103/PhysRevD.83.105019} {\bibfield
  {journal} {\bibinfo  {journal} {Phys. Rev.}\ }\textbf {\bibinfo {volume}
  {D83}},\ \bibinfo {pages} {105019} (\bibinfo {year} {2011})},\ \bibinfo
  {note} {[Erratum: Phys. Rev.D84,069902(2011)]},\ \Eprint
  {http://arxiv.org/abs/1101.4651} {arXiv:1101.4651 [hep-ph]} \BibitemShut
  {NoStop}%
\bibitem [{\citenamefont {Krouppa}\ \emph {et~al.}(2015)\citenamefont
  {Krouppa}, \citenamefont {Ryblewski},\ and\ \citenamefont
  {Strickland}}]{krouppa20151}%
  \BibitemOpen
  \bibfield  {author} {\bibinfo {author} {\bibfnamefont {B.}~\bibnamefont
  {Krouppa}}, \bibinfo {author} {\bibfnamefont {R.}~\bibnamefont {Ryblewski}},
  \ and\ \bibinfo {author} {\bibfnamefont {M.}~\bibnamefont {Strickland}},\
  }\href {\doibase 10.1103/PhysRevC.92.061901} {\bibfield  {journal} {\bibinfo
  {journal} {Phys. Rev.}\ }\textbf {\bibinfo {volume} {C92}},\ \bibinfo {pages}
  {061901} (\bibinfo {year} {2015})},\ \Eprint
  {http://arxiv.org/abs/1507.03951} {arXiv:1507.03951 [hep-ph]} \BibitemShut
  {NoStop}%
\bibitem [{\citenamefont {Krouppa}\ \emph {et~al.}(2018)\citenamefont
  {Krouppa}, \citenamefont {Rothkopf},\ and\ \citenamefont
  {Strickland}}]{krouppa20171}%
  \BibitemOpen
  \bibfield  {author} {\bibinfo {author} {\bibfnamefont {B.}~\bibnamefont
  {Krouppa}}, \bibinfo {author} {\bibfnamefont {A.}~\bibnamefont {Rothkopf}}, \
  and\ \bibinfo {author} {\bibfnamefont {M.}~\bibnamefont {Strickland}},\
  }\href {\doibase 10.1103/PhysRevD.97.016017} {\bibfield  {journal} {\bibinfo
  {journal} {Phys. Rev.}\ }\textbf {\bibinfo {volume} {D97}},\ \bibinfo {pages}
  {016017} (\bibinfo {year} {2018})},\ \Eprint
  {http://arxiv.org/abs/1710.02319} {arXiv:1710.02319 [hep-ph]} \BibitemShut
  {NoStop}%
\bibitem [{\citenamefont {Krouppa}\ \emph {et~al.}(2019)\citenamefont
  {Krouppa}, \citenamefont {Rothkopf},\ and\ \citenamefont
  {Strickland}}]{krouppa20181}%
  \BibitemOpen
  \bibfield  {author} {\bibinfo {author} {\bibfnamefont {B.}~\bibnamefont
  {Krouppa}}, \bibinfo {author} {\bibfnamefont {A.}~\bibnamefont {Rothkopf}}, \
  and\ \bibinfo {author} {\bibfnamefont {M.}~\bibnamefont {Strickland}},\
  }\bibfield  {booktitle} {\emph {\bibinfo {booktitle} {{Proceedings, 27th
  International Conference on Ultrarelativistic Nucleus-Nucleus Collisions
  (Quark Matter 2018): Venice, Italy, May 14-19, 2018}}},\ }\href {\doibase
  10.1016/j.nuclphysa.2018.09.034} {\bibfield  {journal} {\bibinfo  {journal}
  {Nucl. Phys.}\ }\textbf {\bibinfo {volume} {A982}},\ \bibinfo {pages} {727}
  (\bibinfo {year} {2019})},\ \Eprint {http://arxiv.org/abs/1807.07452}
  {arXiv:1807.07452 [hep-ph]} \BibitemShut {NoStop}%
\bibitem [{\citenamefont {{R. Katz and P. B. Gossiaux}}(2016)}]{Katz:2015qja}%
  \BibitemOpen
  \bibfield  {author} {\bibinfo {author} {\bibnamefont {{R. Katz and P. B.
  Gossiaux}}},\ }\href {\doibase 10.1016/j.aop.2016.02.005} {\bibfield
  {journal} {\bibinfo  {journal} {Annals Phys.}\ }\textbf {\bibinfo {volume}
  {368}},\ \bibinfo {pages} {267} (\bibinfo {year} {2016})},\ \Eprint
  {http://arxiv.org/abs/1504.08087} {arXiv:1504.08087 [quant-ph]} \BibitemShut
  {NoStop}%
\bibitem [{\citenamefont {{P. B. Gossiaux and R.
  Katz}}(2016)}]{Gossiaux:2016htk}%
  \BibitemOpen
  \bibfield  {author} {\bibinfo {author} {\bibnamefont {{P. B. Gossiaux and R.
  Katz}}},\ }\bibfield  {booktitle} {\emph {\bibinfo {booktitle} {{Proceedings,
  25th International Conference on Ultra-Relativistic Nucleus-Nucleus
  Collisions (Quark Matter 2015): Kobe, Japan, September 27-October 3,
  2015}}},\ }\href {\doibase 10.1016/j.nuclphysa.2016.04.017} {\bibfield
  {journal} {\bibinfo  {journal} {Nucl. Phys.}\ }\textbf {\bibinfo {volume}
  {A956}},\ \bibinfo {pages} {737} (\bibinfo {year} {2016})},\ \Eprint
  {http://arxiv.org/abs/1601.01443} {arXiv:1601.01443 [hep-ph]} \BibitemShut
  {NoStop}%
\bibitem [{\citenamefont {{P. B. Gossiaux and R.
  Katz}}(2017)}]{Bernard:2016spw}%
  \BibitemOpen
  \bibfield  {author} {\bibinfo {author} {\bibnamefont {{P. B. Gossiaux and R.
  Katz}}},\ }\bibfield  {booktitle} {\emph {\bibinfo {booktitle} {{Proceedings,
  16th International Conference on Strangeness in Quark Matter (SQM 2016):
  Berkeley, California, United States}}},\ }\href {\doibase
  10.1088/1742-6596/779/1/012041} {\bibfield  {journal} {\bibinfo  {journal}
  {J. Phys. Conf. Ser.}\ }\textbf {\bibinfo {volume} {779}},\ \bibinfo {pages}
  {012041} (\bibinfo {year} {2017})},\ \Eprint
  {http://arxiv.org/abs/1611.06499} {arXiv:1611.06499 [hep-ph]} \BibitemShut
  {NoStop}%
\bibitem [{\citenamefont {{R. Sharma and I. Vitev}}(2013)}]{sharma20121}%
  \BibitemOpen
  \bibfield  {author} {\bibinfo {author} {\bibnamefont {{R. Sharma and I.
  Vitev}}},\ }\href {\doibase 10.1103/PhysRevC.87.044905} {\bibfield  {journal}
  {\bibinfo  {journal} {Phys. Rev.}\ }\textbf {\bibinfo {volume} {C87}},\
  \bibinfo {pages} {044905} (\bibinfo {year} {2013})},\ \Eprint
  {http://arxiv.org/abs/1203.0329} {arXiv:1203.0329 [hep-ph]} \BibitemShut
  {NoStop}%
\bibitem [{\citenamefont {Aronson}\ \emph {et~al.}(2018)\citenamefont
  {Aronson}, \citenamefont {Borras}, \citenamefont {Odegard}, \citenamefont
  {Sharma},\ and\ \citenamefont {Vitev}}]{aronson20171}%
  \BibitemOpen
  \bibfield  {author} {\bibinfo {author} {\bibfnamefont {S.}~\bibnamefont
  {Aronson}}, \bibinfo {author} {\bibfnamefont {E.}~\bibnamefont {Borras}},
  \bibinfo {author} {\bibfnamefont {B.}~\bibnamefont {Odegard}}, \bibinfo
  {author} {\bibfnamefont {R.}~\bibnamefont {Sharma}}, \ and\ \bibinfo {author}
  {\bibfnamefont {I.}~\bibnamefont {Vitev}},\ }\href {\doibase
  10.1016/j.physletb.2018.01.038} {\bibfield  {journal} {\bibinfo  {journal}
  {Phys. Lett.}\ }\textbf {\bibinfo {volume} {B778}},\ \bibinfo {pages} {384}
  (\bibinfo {year} {2018})},\ \Eprint {http://arxiv.org/abs/1709.02372}
  {arXiv:1709.02372 [hep-ph]} \BibitemShut {NoStop}%
\bibitem [{\citenamefont {{Y. Makris and I. Vitev}}(2019)}]{makris20191}%
  \BibitemOpen
  \bibfield  {author} {\bibinfo {author} {\bibnamefont {{Y. Makris and I.
  Vitev}}},\ }\href@noop {} {\  (\bibinfo {year} {2019})},\ \Eprint
  {http://arxiv.org/abs/1906.04186} {arXiv:1906.04186 [hep-ph]} \BibitemShut
  {NoStop}%
\bibitem [{\citenamefont {Kajimoto}\ \emph {et~al.}(2018)\citenamefont
  {Kajimoto}, \citenamefont {Akamatsu}, \citenamefont {Asakawa},\ and\
  \citenamefont {Rothkopf}}]{akamatsu20181}%
  \BibitemOpen
  \bibfield  {author} {\bibinfo {author} {\bibfnamefont {S.}~\bibnamefont
  {Kajimoto}}, \bibinfo {author} {\bibfnamefont {Y.}~\bibnamefont {Akamatsu}},
  \bibinfo {author} {\bibfnamefont {M.}~\bibnamefont {Asakawa}}, \ and\
  \bibinfo {author} {\bibfnamefont {A.}~\bibnamefont {Rothkopf}},\ }\href@noop
  {} {\bibfield  {journal} {\bibinfo  {journal} {Physical Review D}\ }\textbf
  {\bibinfo {volume} {97}},\ \bibinfo {pages} {014003} (\bibinfo {year}
  {2018})}\BibitemShut {NoStop}%
\bibitem [{\citenamefont {Brambilla}\ \emph {et~al.}(2017)\citenamefont
  {Brambilla}, \citenamefont {Escobedo}, \citenamefont {Soto},\ and\
  \citenamefont {Vairo}}]{brambilla20171}%
  \BibitemOpen
  \bibfield  {author} {\bibinfo {author} {\bibfnamefont {N.}~\bibnamefont
  {Brambilla}}, \bibinfo {author} {\bibfnamefont {M.~A.}\ \bibnamefont
  {Escobedo}}, \bibinfo {author} {\bibfnamefont {J.}~\bibnamefont {Soto}}, \
  and\ \bibinfo {author} {\bibfnamefont {A.}~\bibnamefont {Vairo}},\ }\href
  {\doibase 10.1103/PhysRevD.96.034021} {\bibfield  {journal} {\bibinfo
  {journal} {Phys. Rev. D}\ }\textbf {\bibinfo {volume} {96}},\ \bibinfo
  {pages} {034021} (\bibinfo {year} {2017})}\BibitemShut {NoStop}%
\bibitem [{\citenamefont {Brambilla}\ \emph {et~al.}(2018)\citenamefont
  {Brambilla}, \citenamefont {Escobedo}, \citenamefont {Soto},\ and\
  \citenamefont {Vairo}}]{brambilla20181}%
  \BibitemOpen
  \bibfield  {author} {\bibinfo {author} {\bibfnamefont {N.}~\bibnamefont
  {Brambilla}}, \bibinfo {author} {\bibfnamefont {M.~A.}\ \bibnamefont
  {Escobedo}}, \bibinfo {author} {\bibfnamefont {J.}~\bibnamefont {Soto}}, \
  and\ \bibinfo {author} {\bibfnamefont {A.}~\bibnamefont {Vairo}},\ }\href
  {\doibase 10.1103/PhysRevD.97.074009} {\bibfield  {journal} {\bibinfo
  {journal} {Phys. Rev. D}\ }\textbf {\bibinfo {volume} {97}},\ \bibinfo
  {pages} {074009} (\bibinfo {year} {2018})}\BibitemShut {NoStop}%
\bibitem [{\citenamefont {Islam}\ and\ \citenamefont
  {Strickland}(2020)}]{Islam:2020bnp}%
  \BibitemOpen
  \bibfield  {author} {\bibinfo {author} {\bibfnamefont {A.}~\bibnamefont
  {Islam}}\ and\ \bibinfo {author} {\bibfnamefont {M.}~\bibnamefont
  {Strickland}},\ }\href {\doibase 10.1007/JHEP03(2021)235} {\bibfield
  {journal} {\bibinfo  {journal} {JHEP}\ }\textbf {\bibinfo {volume} {21}},\
  \bibinfo {pages} {235} (\bibinfo {year} {2020})},\ \Eprint
  {http://arxiv.org/abs/2010.05457} {arXiv:2010.05457 [hep-ph]} \BibitemShut
  {NoStop}%
\bibitem [{\citenamefont {Sharma}\ and\ \citenamefont
  {Tiwari}(2020)}]{Sharma:2019xum}%
  \BibitemOpen
  \bibfield  {author} {\bibinfo {author} {\bibfnamefont {R.}~\bibnamefont
  {Sharma}}\ and\ \bibinfo {author} {\bibfnamefont {A.}~\bibnamefont
  {Tiwari}},\ }\href {\doibase 10.1103/PhysRevD.101.074004} {\bibfield
  {journal} {\bibinfo  {journal} {Phys. Rev. D}\ }\textbf {\bibinfo {volume}
  {101}},\ \bibinfo {pages} {074004} (\bibinfo {year} {2020})},\ \Eprint
  {http://arxiv.org/abs/1912.07036} {arXiv:1912.07036 [hep-ph]} \BibitemShut
  {NoStop}%
\bibitem [{\citenamefont {Moore}\ and\ \citenamefont
  {Teaney}(2005)}]{moore20051}%
  \BibitemOpen
  \bibfield  {author} {\bibinfo {author} {\bibfnamefont {G.~D.}\ \bibnamefont
  {Moore}}\ and\ \bibinfo {author} {\bibfnamefont {D.}~\bibnamefont {Teaney}},\
  }\href {\doibase 10.1103/PhysRevC.71.064904} {\bibfield  {journal} {\bibinfo
  {journal} {Phys. Rev. C}\ }\textbf {\bibinfo {volume} {71}},\ \bibinfo
  {pages} {064904} (\bibinfo {year} {2005})}\BibitemShut {NoStop}%
\bibitem [{\citenamefont {{A. Mocsy and P. Petreczky}}(2007)}]{ymocs20071}%
  \BibitemOpen
  \bibfield  {author} {\bibinfo {author} {\bibnamefont {{A. Mocsy and P.
  Petreczky}}},\ }\href {\doibase 10.1103/PhysRevLett.99.211602} {\bibfield
  {journal} {\bibinfo  {journal} {Phys. Rev. Lett.}\ }\textbf {\bibinfo
  {volume} {99}},\ \bibinfo {pages} {211602} (\bibinfo {year} {2007})},\
  \Eprint {http://arxiv.org/abs/0706.2183} {arXiv:0706.2183 [hep-ph]}
  \BibitemShut {NoStop}%
\bibitem [{\citenamefont {Brambilla}\ \emph {et~al.}(2010)\citenamefont
  {Brambilla}, \citenamefont {Escobedo}, \citenamefont {Ghiglieri},
  \citenamefont {Soto},\ and\ \citenamefont {Vairo}}]{brambilla20101}%
  \BibitemOpen
  \bibfield  {author} {\bibinfo {author} {\bibfnamefont {N.}~\bibnamefont
  {Brambilla}}, \bibinfo {author} {\bibfnamefont {M.~{\'A}.}\ \bibnamefont
  {Escobedo}}, \bibinfo {author} {\bibfnamefont {J.}~\bibnamefont {Ghiglieri}},
  \bibinfo {author} {\bibfnamefont {J.}~\bibnamefont {Soto}}, \ and\ \bibinfo
  {author} {\bibfnamefont {A.}~\bibnamefont {Vairo}},\ }\href {\doibase
  10.1007/JHEP09(2010)038} {\bibfield  {journal} {\bibinfo  {journal} {Journal
  of High Energy Physics}\ }\textbf {\bibinfo {volume} {2010}},\ \bibinfo
  {pages} {38} (\bibinfo {year} {2010})}\BibitemShut {NoStop}%
\bibitem [{\citenamefont {Brambilla}\ \emph {et~al.}(2011)\citenamefont
  {Brambilla}, \citenamefont {Escobedo}, \citenamefont {Ghiglieri},\ and\
  \citenamefont {Vairo}}]{brambilla20111}%
  \BibitemOpen
  \bibfield  {author} {\bibinfo {author} {\bibfnamefont {N.}~\bibnamefont
  {Brambilla}}, \bibinfo {author} {\bibfnamefont {M.~{\'A}.}\ \bibnamefont
  {Escobedo}}, \bibinfo {author} {\bibfnamefont {J.}~\bibnamefont {Ghiglieri}},
  \ and\ \bibinfo {author} {\bibfnamefont {A.}~\bibnamefont {Vairo}},\ }\href
  {\doibase 10.1007/JHEP12(2011)116} {\bibfield  {journal} {\bibinfo  {journal}
  {Journal of High Energy Physics}\ }\textbf {\bibinfo {volume} {2011}},\
  \bibinfo {pages} {116} (\bibinfo {year} {2011})}\BibitemShut {NoStop}%
\bibitem [{\citenamefont {Brambilla}\ \emph {et~al.}(2013)\citenamefont
  {Brambilla}, \citenamefont {Escobedo}, \citenamefont {Ghiglieri},\ and\
  \citenamefont {Vairo}}]{brambilla20131}%
  \BibitemOpen
  \bibfield  {author} {\bibinfo {author} {\bibfnamefont {N.}~\bibnamefont
  {Brambilla}}, \bibinfo {author} {\bibfnamefont {M.~A.}\ \bibnamefont
  {Escobedo}}, \bibinfo {author} {\bibfnamefont {J.}~\bibnamefont {Ghiglieri}},
  \ and\ \bibinfo {author} {\bibfnamefont {A.}~\bibnamefont {Vairo}},\ }\href
  {\doibase 10.1007/JHEP05(2013)130} {\bibfield  {journal} {\bibinfo  {journal}
  {JHEP}\ }\textbf {\bibinfo {volume} {05}},\ \bibinfo {pages} {130} (\bibinfo
  {year} {2013})},\ \Eprint {http://arxiv.org/abs/1303.6097} {arXiv:1303.6097
  [hep-ph]} \BibitemShut {NoStop}%
\bibitem [{\citenamefont {Bala}\ and\ \citenamefont
  {Datta}(2021)}]{Bala:2020tdt}%
  \BibitemOpen
  \bibfield  {author} {\bibinfo {author} {\bibfnamefont {D.}~\bibnamefont
  {Bala}}\ and\ \bibinfo {author} {\bibfnamefont {S.}~\bibnamefont {Datta}},\
  }\href {\doibase 10.1103/PhysRevD.103.014512} {\bibfield  {journal} {\bibinfo
   {journal} {Phys. Rev. D}\ }\textbf {\bibinfo {volume} {103}},\ \bibinfo
  {pages} {014512} (\bibinfo {year} {2021})},\ \Eprint
  {http://arxiv.org/abs/2009.00773} {arXiv:2009.00773 [hep-lat]} \BibitemShut
  {NoStop}%
\bibitem [{\citenamefont {Bodwin}\ \emph {et~al.}(1995)\citenamefont {Bodwin},
  \citenamefont {Braaten},\ and\ \citenamefont {Lepage}}]{bodwin19941}%
  \BibitemOpen
  \bibfield  {author} {\bibinfo {author} {\bibfnamefont {G.~T.}\ \bibnamefont
  {Bodwin}}, \bibinfo {author} {\bibfnamefont {E.}~\bibnamefont {Braaten}}, \
  and\ \bibinfo {author} {\bibfnamefont {G.~P.}\ \bibnamefont {Lepage}},\
  }\href {\doibase 10.1103/PhysRevD.55.5853, 10.1103/PhysRevD.51.1125}
  {\bibfield  {journal} {\bibinfo  {journal} {Phys. Rev.}\ }\textbf {\bibinfo
  {volume} {D51}},\ \bibinfo {pages} {1125} (\bibinfo {year} {1995})},\
  \bibinfo {note} {[Erratum: Phys. Rev.D55,5853(1997)]},\ \Eprint
  {http://arxiv.org/abs/hep-ph/9407339} {arXiv:hep-ph/9407339 [hep-ph]}
  \BibitemShut {NoStop}%
\bibitem [{\citenamefont {Akamatsu}(2013)}]{akamatsu20131}%
  \BibitemOpen
  \bibfield  {author} {\bibinfo {author} {\bibfnamefont {Y.}~\bibnamefont
  {Akamatsu}},\ }\href@noop {} {\bibfield  {journal} {\bibinfo  {journal}
  {Physical Review D}\ }\textbf {\bibinfo {volume} {87}},\ \bibinfo {pages}
  {045016} (\bibinfo {year} {2013})}\BibitemShut {NoStop}%
\bibitem [{\citenamefont {Rothkopf}\ \emph {et~al.}(2012)\citenamefont
  {Rothkopf}, \citenamefont {Hatsuda},\ and\ \citenamefont
  {Sasaki}}]{rothkopf20111}%
  \BibitemOpen
  \bibfield  {author} {\bibinfo {author} {\bibfnamefont {A.}~\bibnamefont
  {Rothkopf}}, \bibinfo {author} {\bibfnamefont {T.}~\bibnamefont {Hatsuda}}, \
  and\ \bibinfo {author} {\bibfnamefont {S.}~\bibnamefont {Sasaki}},\ }\href
  {\doibase 10.1103/PhysRevLett.108.162001} {\bibfield  {journal} {\bibinfo
  {journal} {Phys. Rev. Lett.}\ }\textbf {\bibinfo {volume} {108}},\ \bibinfo
  {pages} {162001} (\bibinfo {year} {2012})},\ \Eprint
  {http://arxiv.org/abs/1108.1579} {arXiv:1108.1579 [hep-lat]} \BibitemShut
  {NoStop}%
\bibitem [{\citenamefont {Burnier}\ \emph {et~al.}(2015)\citenamefont
  {Burnier}, \citenamefont {Kaczmarek},\ and\ \citenamefont
  {Rothkopf}}]{Burnier:2014ssa}%
  \BibitemOpen
  \bibfield  {author} {\bibinfo {author} {\bibfnamefont {Y.}~\bibnamefont
  {Burnier}}, \bibinfo {author} {\bibfnamefont {O.}~\bibnamefont {Kaczmarek}},
  \ and\ \bibinfo {author} {\bibfnamefont {A.}~\bibnamefont {Rothkopf}},\
  }\href {\doibase 10.1103/PhysRevLett.114.082001} {\bibfield  {journal}
  {\bibinfo  {journal} {Phys. Rev. Lett.}\ }\textbf {\bibinfo {volume} {114}},\
  \bibinfo {pages} {082001} (\bibinfo {year} {2015})},\ \Eprint
  {http://arxiv.org/abs/1410.2546} {arXiv:1410.2546 [hep-lat]} \BibitemShut
  {NoStop}%
\bibitem [{\citenamefont {Petreczky}(2012)}]{petreczky20121}%
  \BibitemOpen
  \bibfield  {author} {\bibinfo {author} {\bibfnamefont {P.}~\bibnamefont
  {Petreczky}},\ }\href {\doibase 10.1088/0954-3899/39/9/093002} {\bibfield
  {journal} {\bibinfo  {journal} {Journal of Physics G: Nuclear and Particle
  Physics}\ }\textbf {\bibinfo {volume} {39}},\ \bibinfo {pages} {093002}
  (\bibinfo {year} {2012})}\BibitemShut {NoStop}%
\bibitem [{\citenamefont {Bala}\ and\ \citenamefont
  {Datta}(2020)}]{Bala:2019cqu}%
  \BibitemOpen
  \bibfield  {author} {\bibinfo {author} {\bibfnamefont {D.}~\bibnamefont
  {Bala}}\ and\ \bibinfo {author} {\bibfnamefont {S.}~\bibnamefont {Datta}},\
  }\href {\doibase 10.1103/PhysRevD.101.034507} {\bibfield  {journal} {\bibinfo
   {journal} {Phys. Rev. D}\ }\textbf {\bibinfo {volume} {101}},\ \bibinfo
  {pages} {034507} (\bibinfo {year} {2020})},\ \Eprint
  {http://arxiv.org/abs/1909.10548} {arXiv:1909.10548 [hep-lat]} \BibitemShut
  {NoStop}%
\bibitem [{\citenamefont {Bala}\ \emph {et~al.}(2022)\citenamefont {Bala},
  \citenamefont {Kaczmarek}, \citenamefont {Larsen}, \citenamefont {Mukherjee},
  \citenamefont {Parkar}, \citenamefont {Petreczky}, \citenamefont {Rothkopf},\
  and\ \citenamefont {Weber}}]{Bala:2021fkm}%
  \BibitemOpen
  \bibfield  {author} {\bibinfo {author} {\bibfnamefont {D.}~\bibnamefont
  {Bala}}, \bibinfo {author} {\bibfnamefont {O.}~\bibnamefont {Kaczmarek}},
  \bibinfo {author} {\bibfnamefont {R.}~\bibnamefont {Larsen}}, \bibinfo
  {author} {\bibfnamefont {S.}~\bibnamefont {Mukherjee}}, \bibinfo {author}
  {\bibfnamefont {G.}~\bibnamefont {Parkar}}, \bibinfo {author} {\bibfnamefont
  {P.}~\bibnamefont {Petreczky}}, \bibinfo {author} {\bibfnamefont
  {A.}~\bibnamefont {Rothkopf}}, \ and\ \bibinfo {author} {\bibfnamefont
  {J.~H.}\ \bibnamefont {Weber}} (\bibinfo {collaboration} {HotQCD}),\ }\href
  {\doibase 10.1103/PhysRevD.105.054513} {\bibfield  {journal} {\bibinfo
  {journal} {Phys. Rev. D}\ }\textbf {\bibinfo {volume} {105}},\ \bibinfo
  {pages} {054513} (\bibinfo {year} {2022})},\ \Eprint
  {http://arxiv.org/abs/2110.11659} {arXiv:2110.11659 [hep-lat]} \BibitemShut
  {NoStop}%
\bibitem [{\citenamefont {Miura}\ \emph {et~al.}(2019)\citenamefont {Miura},
  \citenamefont {Akamatsu}, \citenamefont {Asakawa},\ and\ \citenamefont
  {Rothkopf}}]{akamatsu20191}%
  \BibitemOpen
  \bibfield  {author} {\bibinfo {author} {\bibfnamefont {T.}~\bibnamefont
  {Miura}}, \bibinfo {author} {\bibfnamefont {Y.}~\bibnamefont {Akamatsu}},
  \bibinfo {author} {\bibfnamefont {M.}~\bibnamefont {Asakawa}}, \ and\
  \bibinfo {author} {\bibfnamefont {A.}~\bibnamefont {Rothkopf}},\ }\href@noop
  {} {\  (\bibinfo {year} {2019})},\ \Eprint {http://arxiv.org/abs/1908.06293}
  {arXiv:1908.06293 [nucl-th]} \BibitemShut {NoStop}%
\bibitem [{\citenamefont {Kobes}(1991)}]{Kobes:1990ua}%
  \BibitemOpen
  \bibfield  {author} {\bibinfo {author} {\bibfnamefont {R.}~\bibnamefont
  {Kobes}},\ }\href {\doibase 10.1103/PhysRevD.43.1269} {\bibfield  {journal}
  {\bibinfo  {journal} {Phys. Rev. D}\ }\textbf {\bibinfo {volume} {43}},\
  \bibinfo {pages} {1269} (\bibinfo {year} {1991})}\BibitemShut {NoStop}%
\bibitem [{\citenamefont {Kobes}\ and\ \citenamefont
  {Semenoff}(1986)}]{Kobes:1986za}%
  \BibitemOpen
  \bibfield  {author} {\bibinfo {author} {\bibfnamefont {R.~L.}\ \bibnamefont
  {Kobes}}\ and\ \bibinfo {author} {\bibfnamefont {G.~W.}\ \bibnamefont
  {Semenoff}},\ }\href {\doibase 10.1016/0550-3213(86)90006-4} {\bibfield
  {journal} {\bibinfo  {journal} {Nucl. Phys. B}\ }\textbf {\bibinfo {volume}
  {272}},\ \bibinfo {pages} {329} (\bibinfo {year} {1986})}\BibitemShut
  {NoStop}%
\bibitem [{\citenamefont {Kapusta}\ and\ \citenamefont
  {Gale}(2006)}]{kapusta_gale_2006}%
  \BibitemOpen
  \bibfield  {author} {\bibinfo {author} {\bibfnamefont {J.~I.}\ \bibnamefont
  {Kapusta}}\ and\ \bibinfo {author} {\bibfnamefont {C.}~\bibnamefont {Gale}},\
  }\href {\doibase 10.1017/CBO9780511535130} {\emph {\bibinfo {title}
  {Finite-Temperature Field Theory: Principles and Applications}}},\ \bibinfo
  {edition} {2nd}\ ed.,\ Cambridge Monographs on Mathematical Physics\
  (\bibinfo  {publisher} {Cambridge University Press},\ \bibinfo {year}
  {2006})\BibitemShut {NoStop}%
\bibitem [{\citenamefont {Bellac}(2011)}]{Bellac:2011kqa}%
  \BibitemOpen
  \bibfield  {author} {\bibinfo {author} {\bibfnamefont {M.~L.}\ \bibnamefont
  {Bellac}},\ }\href {\doibase 10.1017/CBO9780511721700} {\emph {\bibinfo
  {title} {{Thermal Field Theory}}}},\ Cambridge Monographs on Mathematical
  Physics\ (\bibinfo  {publisher} {Cambridge University Press},\ \bibinfo
  {year} {2011})\BibitemShut {NoStop}%
\bibitem [{\citenamefont {{J. Casalderrey-Solana and D.
  Teaney}}(2006)}]{solana20061}%
  \BibitemOpen
  \bibfield  {author} {\bibinfo {author} {\bibnamefont {{J. Casalderrey-Solana
  and D. Teaney}}},\ }\href {\doibase 10.1103/PhysRevD.74.085012} {\bibfield
  {journal} {\bibinfo  {journal} {Phys. Rev.}\ }\textbf {\bibinfo {volume}
  {D74}},\ \bibinfo {pages} {085012} (\bibinfo {year} {2006})},\ \Eprint
  {http://arxiv.org/abs/hep-ph/0605199} {arXiv:hep-ph/0605199 [hep-ph]}
  \BibitemShut {NoStop}%
\bibitem [{\citenamefont {{C. H. Simon and G. D. Moore}}(2008)}]{huot20081}%
  \BibitemOpen
  \bibfield  {author} {\bibinfo {author} {\bibnamefont {{C. H. Simon and G. D.
  Moore}}},\ }\href {\doibase 10.1088/1126-6708/2008/02/081} {\bibfield
  {journal} {\bibinfo  {journal} {Journal of High Energy Physics}\ }\textbf
  {\bibinfo {volume} {2008}},\ \bibinfo {pages} {081} (\bibinfo {year}
  {2008})}\BibitemShut {NoStop}%
\bibitem [{\citenamefont {Francis}\ \emph
  {et~al.}(2015{\natexlab{a}})\citenamefont {Francis}, \citenamefont
  {Kaczmarek}, \citenamefont {Laine}, \citenamefont {Neuhaus},\ and\
  \citenamefont {Ohno}}]{Francis:2015daa}%
  \BibitemOpen
  \bibfield  {author} {\bibinfo {author} {\bibfnamefont {A.}~\bibnamefont
  {Francis}}, \bibinfo {author} {\bibfnamefont {O.}~\bibnamefont {Kaczmarek}},
  \bibinfo {author} {\bibfnamefont {M.}~\bibnamefont {Laine}}, \bibinfo
  {author} {\bibfnamefont {T.}~\bibnamefont {Neuhaus}}, \ and\ \bibinfo
  {author} {\bibfnamefont {H.}~\bibnamefont {Ohno}},\ }\href {\doibase
  10.1103/PhysRevD.92.116003} {\bibfield  {journal} {\bibinfo  {journal} {Phys.
  Rev. D}\ }\textbf {\bibinfo {volume} {92}},\ \bibinfo {pages} {116003}
  (\bibinfo {year} {2015}{\natexlab{a}})},\ \Eprint
  {http://arxiv.org/abs/1508.04543} {arXiv:1508.04543 [hep-lat]} \BibitemShut
  {NoStop}%
\bibitem [{\citenamefont {Eller}\ \emph {et~al.}(2019)\citenamefont {Eller},
  \citenamefont {Ghiglieri},\ and\ \citenamefont {Moore}}]{Eller:2019spw}%
  \BibitemOpen
  \bibfield  {author} {\bibinfo {author} {\bibfnamefont {A.~M.}\ \bibnamefont
  {Eller}}, \bibinfo {author} {\bibfnamefont {J.}~\bibnamefont {Ghiglieri}}, \
  and\ \bibinfo {author} {\bibfnamefont {G.~D.}\ \bibnamefont {Moore}},\ }\href
  {\doibase 10.1103/PhysRevD.99.094042} {\bibfield  {journal} {\bibinfo
  {journal} {Phys. Rev. D}\ }\textbf {\bibinfo {volume} {99}},\ \bibinfo
  {pages} {094042} (\bibinfo {year} {2019})},\ \bibinfo {note} {[Erratum:
  Phys.Rev.D 102, 039901 (2020)]},\ \Eprint {http://arxiv.org/abs/1903.08064}
  {arXiv:1903.08064 [hep-ph]} \BibitemShut {NoStop}%
\bibitem [{\citenamefont {Banerjee}\ \emph {et~al.}(2012)\citenamefont
  {Banerjee}, \citenamefont {Datta}, \citenamefont {Gavai},\ and\ \citenamefont
  {Majumdar}}]{datta20121}%
  \BibitemOpen
  \bibfield  {author} {\bibinfo {author} {\bibfnamefont {D.}~\bibnamefont
  {Banerjee}}, \bibinfo {author} {\bibfnamefont {S.}~\bibnamefont {Datta}},
  \bibinfo {author} {\bibfnamefont {R.}~\bibnamefont {Gavai}}, \ and\ \bibinfo
  {author} {\bibfnamefont {P.}~\bibnamefont {Majumdar}},\ }\href {\doibase
  10.1103/PhysRevD.85.014510} {\bibfield  {journal} {\bibinfo  {journal} {Phys.
  Rev. D}\ }\textbf {\bibinfo {volume} {85}},\ \bibinfo {pages} {014510}
  (\bibinfo {year} {2012})}\BibitemShut {NoStop}%
\bibitem [{\citenamefont {Francis}\ \emph
  {et~al.}(2015{\natexlab{b}})\citenamefont {Francis}, \citenamefont
  {Kaczmarek}, \citenamefont {Laine}, \citenamefont {Neuhaus},\ and\
  \citenamefont {Ohno}}]{laine20151}%
  \BibitemOpen
  \bibfield  {author} {\bibinfo {author} {\bibfnamefont {A.}~\bibnamefont
  {Francis}}, \bibinfo {author} {\bibfnamefont {O.}~\bibnamefont {Kaczmarek}},
  \bibinfo {author} {\bibfnamefont {M.}~\bibnamefont {Laine}}, \bibinfo
  {author} {\bibfnamefont {T.}~\bibnamefont {Neuhaus}}, \ and\ \bibinfo
  {author} {\bibfnamefont {H.}~\bibnamefont {Ohno}},\ }\href {\doibase
  10.1103/PhysRevD.92.116003} {\bibfield  {journal} {\bibinfo  {journal} {Phys.
  Rev. D}\ }\textbf {\bibinfo {volume} {92}},\ \bibinfo {pages} {116003}
  (\bibinfo {year} {2015}{\natexlab{b}})}\BibitemShut {NoStop}%
\bibitem [{\citenamefont {Banerjee}\ \emph {et~al.}(2022)\citenamefont
  {Banerjee}, \citenamefont {Gavai}, \citenamefont {Datta},\ and\ \citenamefont
  {Majumdar}}]{Banerjee:2022gen}%
  \BibitemOpen
  \bibfield  {author} {\bibinfo {author} {\bibfnamefont {D.}~\bibnamefont
  {Banerjee}}, \bibinfo {author} {\bibfnamefont {R.}~\bibnamefont {Gavai}},
  \bibinfo {author} {\bibfnamefont {S.}~\bibnamefont {Datta}}, \ and\ \bibinfo
  {author} {\bibfnamefont {P.}~\bibnamefont {Majumdar}},\ }\href@noop {} {\
  (\bibinfo {year} {2022})},\ \Eprint {http://arxiv.org/abs/2206.15471}
  {arXiv:2206.15471 [hep-ph]} \BibitemShut {NoStop}%
\bibitem [{\citenamefont {Brambilla}\ \emph {et~al.}(2022)\citenamefont
  {Brambilla}, \citenamefont {Leino}, \citenamefont {Mayer-Steudte},\ and\
  \citenamefont {Petreczky}}]{Brambilla:2022xbd}%
  \BibitemOpen
  \bibfield  {author} {\bibinfo {author} {\bibfnamefont {N.}~\bibnamefont
  {Brambilla}}, \bibinfo {author} {\bibfnamefont {V.}~\bibnamefont {Leino}},
  \bibinfo {author} {\bibfnamefont {J.}~\bibnamefont {Mayer-Steudte}}, \ and\
  \bibinfo {author} {\bibfnamefont {P.}~\bibnamefont {Petreczky}} (\bibinfo
  {collaboration} {TUMQCD}),\ }\href@noop {} {\  (\bibinfo {year} {2022})},\
  \Eprint {http://arxiv.org/abs/2206.02861} {arXiv:2206.02861 [hep-lat]}
  \BibitemShut {NoStop}%
\bibitem [{\citenamefont {Brambilla}\ \emph {et~al.}(2019)\citenamefont
  {Brambilla}, \citenamefont {Leino}, \citenamefont {Petreczky},\ and\
  \citenamefont {vairo}}]{Brambilla:2019oaa}%
  \BibitemOpen
  \bibfield  {author} {\bibinfo {author} {\bibfnamefont {N.}~\bibnamefont
  {Brambilla}}, \bibinfo {author} {\bibfnamefont {V.}~\bibnamefont {Leino}},
  \bibinfo {author} {\bibfnamefont {P.}~\bibnamefont {Petreczky}}, \ and\
  \bibinfo {author} {\bibfnamefont {A.}~\bibnamefont {vairo}} (\bibinfo
  {collaboration} {TUMQCD}),\ }in\ \href@noop {} {\emph {\bibinfo {booktitle}
  {{37th International Symposium on Lattice Field Theory (Lattice 2019) Wuhan,
  Hubei, China, June 16-22, 2019}}}}\ (\bibinfo {year} {2019})\ \Eprint
  {http://arxiv.org/abs/1912.00689} {arXiv:1912.00689 [hep-lat]} \BibitemShut
  {NoStop}%
\bibitem [{\citenamefont {Kaczmarek}\ and\ \citenamefont
  {Zantow}(2005)}]{Kaczmarek:2005ui}%
  \BibitemOpen
  \bibfield  {author} {\bibinfo {author} {\bibfnamefont {O.}~\bibnamefont
  {Kaczmarek}}\ and\ \bibinfo {author} {\bibfnamefont {F.}~\bibnamefont
  {Zantow}},\ }\href {\doibase 10.1103/PhysRevD.71.114510} {\bibfield
  {journal} {\bibinfo  {journal} {Phys. Rev. D}\ }\textbf {\bibinfo {volume}
  {71}},\ \bibinfo {pages} {114510} (\bibinfo {year} {2005})},\ \Eprint
  {http://arxiv.org/abs/hep-lat/0503017} {arXiv:hep-lat/0503017} \BibitemShut
  {NoStop}%
\bibitem [{\citenamefont {Park}\ \emph {et~al.}(2007)\citenamefont {Park},
  \citenamefont {Kim}, \citenamefont {Song}, \citenamefont {Lee},\ and\
  \citenamefont {Wong}}]{Park:2007zza}%
  \BibitemOpen
  \bibfield  {author} {\bibinfo {author} {\bibfnamefont {Y.}~\bibnamefont
  {Park}}, \bibinfo {author} {\bibfnamefont {K.-I.}\ \bibnamefont {Kim}},
  \bibinfo {author} {\bibfnamefont {T.}~\bibnamefont {Song}}, \bibinfo {author}
  {\bibfnamefont {S.~H.}\ \bibnamefont {Lee}}, \ and\ \bibinfo {author}
  {\bibfnamefont {C.-Y.}\ \bibnamefont {Wong}},\ }\href {\doibase
  10.1103/PhysRevC.76.044907} {\bibfield  {journal} {\bibinfo  {journal} {Phys.
  Rev. C}\ }\textbf {\bibinfo {volume} {76}},\ \bibinfo {pages} {044907}
  (\bibinfo {year} {2007})},\ \Eprint {http://arxiv.org/abs/0704.3770}
  {arXiv:0704.3770 [hep-ph]} \BibitemShut {NoStop}%
\bibitem [{\citenamefont {Sharma}\ and\ \citenamefont
  {Vitev}(2013)}]{Sharma:2012dy}%
  \BibitemOpen
  \bibfield  {author} {\bibinfo {author} {\bibfnamefont {R.}~\bibnamefont
  {Sharma}}\ and\ \bibinfo {author} {\bibfnamefont {I.}~\bibnamefont {Vitev}},\
  }\href {\doibase 10.1103/PhysRevC.87.044905} {\bibfield  {journal} {\bibinfo
  {journal} {Phys. Rev. C}\ }\textbf {\bibinfo {volume} {87}},\ \bibinfo
  {pages} {044905} (\bibinfo {year} {2013})},\ \Eprint
  {http://arxiv.org/abs/1203.0329} {arXiv:1203.0329 [hep-ph]} \BibitemShut
  {NoStop}%
\bibitem [{\citenamefont {Strickland}(2011)}]{Strickland:2011mw}%
  \BibitemOpen
  \bibfield  {author} {\bibinfo {author} {\bibfnamefont {M.}~\bibnamefont
  {Strickland}},\ }\href {\doibase 10.1103/PhysRevLett.107.132301} {\bibfield
  {journal} {\bibinfo  {journal} {Phys. Rev. Lett.}\ }\textbf {\bibinfo
  {volume} {107}},\ \bibinfo {pages} {132301} (\bibinfo {year} {2011})},\
  \Eprint {http://arxiv.org/abs/1106.2571} {arXiv:1106.2571 [hep-ph]}
  \BibitemShut {NoStop}%
\bibitem [{\citenamefont {Dumitru}\ \emph {et~al.}(2009)\citenamefont
  {Dumitru}, \citenamefont {Guo}, \citenamefont {Mocsy},\ and\ \citenamefont
  {Strickland}}]{Dumitru:2009ni}%
  \BibitemOpen
  \bibfield  {author} {\bibinfo {author} {\bibfnamefont {A.}~\bibnamefont
  {Dumitru}}, \bibinfo {author} {\bibfnamefont {Y.}~\bibnamefont {Guo}},
  \bibinfo {author} {\bibfnamefont {A.}~\bibnamefont {Mocsy}}, \ and\ \bibinfo
  {author} {\bibfnamefont {M.}~\bibnamefont {Strickland}},\ }\href {\doibase
  10.1103/PhysRevD.79.054019} {\bibfield  {journal} {\bibinfo  {journal} {Phys.
  Rev. D}\ }\textbf {\bibinfo {volume} {79}},\ \bibinfo {pages} {054019}
  (\bibinfo {year} {2009})},\ \Eprint {http://arxiv.org/abs/0901.1998}
  {arXiv:0901.1998 [hep-ph]} \BibitemShut {NoStop}%
\bibitem [{\citenamefont {Burnier}\ and\ \citenamefont
  {Rothkopf}(2017)}]{Burnier:2016mxc}%
  \BibitemOpen
  \bibfield  {author} {\bibinfo {author} {\bibfnamefont {Y.}~\bibnamefont
  {Burnier}}\ and\ \bibinfo {author} {\bibfnamefont {A.}~\bibnamefont
  {Rothkopf}},\ }\href {\doibase 10.1103/PhysRevD.95.054511} {\bibfield
  {journal} {\bibinfo  {journal} {Phys. Rev. D}\ }\textbf {\bibinfo {volume}
  {95}},\ \bibinfo {pages} {054511} (\bibinfo {year} {2017})},\ \Eprint
  {http://arxiv.org/abs/1607.04049} {arXiv:1607.04049 [hep-lat]} \BibitemShut
  {NoStop}%
\bibitem [{\citenamefont {Du}\ \emph {et~al.}(2017{\natexlab{b}})\citenamefont
  {Du}, \citenamefont {He},\ and\ \citenamefont {Rapp}}]{Du:2017hss}%
  \BibitemOpen
  \bibfield  {author} {\bibinfo {author} {\bibfnamefont {X.}~\bibnamefont
  {Du}}, \bibinfo {author} {\bibfnamefont {M.}~\bibnamefont {He}}, \ and\
  \bibinfo {author} {\bibfnamefont {R.}~\bibnamefont {Rapp}},\ }\href {\doibase
  10.1016/j.nuclphysa.2017.05.079} {\bibfield  {journal} {\bibinfo  {journal}
  {Nucl. Phys. A}\ }\textbf {\bibinfo {volume} {967}},\ \bibinfo {pages} {904}
  (\bibinfo {year} {2017}{\natexlab{b}})},\ \Eprint
  {http://arxiv.org/abs/1704.04838} {arXiv:1704.04838 [hep-ph]} \BibitemShut
  {NoStop}%
\bibitem [{\citenamefont {CMS}(2022)}]{CMS:2022rna}%
  \BibitemOpen
  \bibfield  {author} {\bibinfo {author} {\bibnamefont {CMS}},\ }\bibfield
  {booktitle} {\emph {\bibinfo {booktitle} {{Observation of the
  $\Upsilon\textrm{(3S)}$ meson and sequential suppression of $\Upsilon$ states
  in PbPb collisions at $\sqrt{\mathrm{s_{NN}}}=5.02~\mathrm{TeV}$}}},\
  }\href@noop {} {\bibfield  {journal} {\bibinfo  {journal}
  {CMS-PAS-HIN-21-007}\ } (\bibinfo {year} {2022})}\BibitemShut {NoStop}%
\bibitem [{\citenamefont {Abramowitz}\ and\ \citenamefont
  {Stegun}(1964)}]{abramowitz+stegun}%
  \BibitemOpen
  \bibfield  {author} {\bibinfo {author} {\bibfnamefont {M.}~\bibnamefont
  {Abramowitz}}\ and\ \bibinfo {author} {\bibfnamefont {I.~A.}\ \bibnamefont
  {Stegun}},\ }\href@noop {} {\emph {\bibinfo {title} {Handbook of Mathematical
  Functions with Formulas, Graphs, and Mathematical Tables}}},\ \bibinfo
  {edition} {ninth dover printing, tenth gpo printing}\ ed.\ (\bibinfo
  {publisher} {Dover},\ \bibinfo {address} {New York},\ \bibinfo {year}
  {1964})\BibitemShut {NoStop}%
\bibitem [{\citenamefont {Bjorken}(1983)}]{Bjorken:1982qr}%
  \BibitemOpen
  \bibfield  {author} {\bibinfo {author} {\bibfnamefont {J.~D.}\ \bibnamefont
  {Bjorken}},\ }\href {\doibase 10.1103/PhysRevD.27.140} {\bibfield  {journal}
  {\bibinfo  {journal} {Phys. Rev. D}\ }\textbf {\bibinfo {volume} {27}},\
  \bibinfo {pages} {140} (\bibinfo {year} {1983})}\BibitemShut {NoStop}%
\bibitem [{\citenamefont {Chang}\ \emph {et~al.}(2016)\citenamefont {Chang}
  \emph {et~al.}}]{Chang:2015hqa}%
  \BibitemOpen
  \bibfield  {author} {\bibinfo {author} {\bibfnamefont {N.-b.}\ \bibnamefont
  {Chang}} \emph {et~al.},\ }\href {\doibase 10.1007/s11433-015-5778-0}
  {\bibfield  {journal} {\bibinfo  {journal} {Sci. China Phys. Mech. Astron.}\
  }\textbf {\bibinfo {volume} {59}},\ \bibinfo {pages} {621001} (\bibinfo
  {year} {2016})},\ \Eprint {http://arxiv.org/abs/1510.05754} {arXiv:1510.05754
  [nucl-th]} \BibitemShut {NoStop}%
\bibitem [{\citenamefont {Adam}\ \emph
  {et~al.}(2016{\natexlab{a}})\citenamefont {Adam} \emph
  {et~al.}}]{ALICE:2015juo}%
  \BibitemOpen
  \bibfield  {author} {\bibinfo {author} {\bibfnamefont {J.}~\bibnamefont
  {Adam}} \emph {et~al.} (\bibinfo {collaboration} {ALICE}),\ }\href {\doibase
  10.1103/PhysRevLett.116.222302} {\bibfield  {journal} {\bibinfo  {journal}
  {Phys. Rev. Lett.}\ }\textbf {\bibinfo {volume} {116}},\ \bibinfo {pages}
  {222302} (\bibinfo {year} {2016}{\natexlab{a}})},\ \Eprint
  {http://arxiv.org/abs/1512.06104} {arXiv:1512.06104 [nucl-ex]} \BibitemShut
  {NoStop}%
\bibitem [{\citenamefont {Miller}\ \emph {et~al.}(2007)\citenamefont {Miller},
  \citenamefont {Reygers}, \citenamefont {Sanders},\ and\ \citenamefont
  {Steinberg}}]{Miller:2007ri}%
  \BibitemOpen
  \bibfield  {author} {\bibinfo {author} {\bibfnamefont {M.~L.}\ \bibnamefont
  {Miller}}, \bibinfo {author} {\bibfnamefont {K.}~\bibnamefont {Reygers}},
  \bibinfo {author} {\bibfnamefont {S.~J.}\ \bibnamefont {Sanders}}, \ and\
  \bibinfo {author} {\bibfnamefont {P.}~\bibnamefont {Steinberg}},\ }\href
  {\doibase 10.1146/annurev.nucl.57.090506.123020} {\bibfield  {journal}
  {\bibinfo  {journal} {Ann. Rev. Nucl. Part. Sci.}\ }\textbf {\bibinfo
  {volume} {57}},\ \bibinfo {pages} {205} (\bibinfo {year} {2007})},\ \Eprint
  {http://arxiv.org/abs/nucl-ex/0701025} {arXiv:nucl-ex/0701025} \BibitemShut
  {NoStop}%
\bibitem [{\citenamefont {Abelev}\ \emph {et~al.}(2013)\citenamefont {Abelev}
  \emph {et~al.}}]{ALICE:2013hur}%
  \BibitemOpen
  \bibfield  {author} {\bibinfo {author} {\bibfnamefont {B.}~\bibnamefont
  {Abelev}} \emph {et~al.} (\bibinfo {collaboration} {ALICE}),\ }\href
  {\doibase 10.1103/PhysRevC.88.044909} {\bibfield  {journal} {\bibinfo
  {journal} {Phys. Rev. C}\ }\textbf {\bibinfo {volume} {88}},\ \bibinfo
  {pages} {044909} (\bibinfo {year} {2013})},\ \Eprint
  {http://arxiv.org/abs/1301.4361} {arXiv:1301.4361 [nucl-ex]} \BibitemShut
  {NoStop}%
\bibitem [{\citenamefont {Srivastava}\ \emph {et~al.}(2018)\citenamefont
  {Srivastava}, \citenamefont {Chatterjee},\ and\ \citenamefont
  {Mustafa}}]{Srivastava:2016hwr}%
  \BibitemOpen
  \bibfield  {author} {\bibinfo {author} {\bibfnamefont {D.~K.}\ \bibnamefont
  {Srivastava}}, \bibinfo {author} {\bibfnamefont {R.}~\bibnamefont
  {Chatterjee}}, \ and\ \bibinfo {author} {\bibfnamefont {M.~G.}\ \bibnamefont
  {Mustafa}},\ }\href {\doibase 10.1088/1361-6471/aa9421} {\bibfield  {journal}
  {\bibinfo  {journal} {J. Phys. G}\ }\textbf {\bibinfo {volume} {45}},\
  \bibinfo {pages} {015103} (\bibinfo {year} {2018})},\ \Eprint
  {http://arxiv.org/abs/1609.06496} {arXiv:1609.06496 [nucl-th]} \BibitemShut
  {NoStop}%
\bibitem [{\citenamefont {Adam}\ \emph
  {et~al.}(2016{\natexlab{b}})\citenamefont {Adam} \emph
  {et~al.}}]{ALICE:2016igk}%
  \BibitemOpen
  \bibfield  {author} {\bibinfo {author} {\bibfnamefont {J.}~\bibnamefont
  {Adam}} \emph {et~al.} (\bibinfo {collaboration} {ALICE}),\ }\href {\doibase
  10.1103/PhysRevC.94.034903} {\bibfield  {journal} {\bibinfo  {journal} {Phys.
  Rev. C}\ }\textbf {\bibinfo {volume} {94}},\ \bibinfo {pages} {034903}
  (\bibinfo {year} {2016}{\natexlab{b}})},\ \Eprint
  {http://arxiv.org/abs/1603.04775} {arXiv:1603.04775 [nucl-ex]} \BibitemShut
  {NoStop}%
\bibitem [{\citenamefont {Eyyubova}\ \emph {et~al.}(2021)\citenamefont
  {Eyyubova}, \citenamefont {Korotkikh}, \citenamefont {Snigirev},\ and\
  \citenamefont {Zabrodin}}]{Eyyubova:2021ngi}%
  \BibitemOpen
  \bibfield  {author} {\bibinfo {author} {\bibfnamefont {G.~K.}\ \bibnamefont
  {Eyyubova}}, \bibinfo {author} {\bibfnamefont {V.~L.}\ \bibnamefont
  {Korotkikh}}, \bibinfo {author} {\bibfnamefont {A.~M.}\ \bibnamefont
  {Snigirev}}, \ and\ \bibinfo {author} {\bibfnamefont {E.~E.}\ \bibnamefont
  {Zabrodin}},\ }\href {\doibase 10.1088/1361-6471/ac1079} {\bibfield
  {journal} {\bibinfo  {journal} {J. Phys. G}\ }\textbf {\bibinfo {volume}
  {48}},\ \bibinfo {pages} {095101} (\bibinfo {year} {2021})},\ \Eprint
  {http://arxiv.org/abs/2107.00521} {arXiv:2107.00521 [nucl-th]} \BibitemShut
  {NoStop}%
\bibitem [{\citenamefont {Mikko~Laine}(2007)}]{laine20072}%
  \BibitemOpen
  \bibfield  {author} {\bibinfo {author} {\bibfnamefont {M.~T.}\ \bibnamefont
  {Mikko~Laine}, \bibfnamefont {Owe~Philipsen}},\ }\href {\doibase
  10.1088/1126-6708/2007/09/066} {\bibfield  {journal} {\bibinfo  {journal}
  {Journal of High Energy Physics}\ }\textbf {\bibinfo {volume} {2007}},\
  \bibinfo {pages} {066} (\bibinfo {year} {2007})}\BibitemShut {NoStop}%
\bibitem [{\citenamefont {Akamatsu}(2015)}]{akamatsu20151}%
  \BibitemOpen
  \bibfield  {author} {\bibinfo {author} {\bibfnamefont {Y.}~\bibnamefont
  {Akamatsu}},\ }\href@noop {} {\bibfield  {journal} {\bibinfo  {journal}
  {Physical Review D}\ }\textbf {\bibinfo {volume} {91}},\ \bibinfo {pages}
  {056002} (\bibinfo {year} {2015})}\BibitemShut {NoStop}%
\bibitem [{\citenamefont {Akamatsu}(2022)}]{Akamatsu:2020ypb}%
  \BibitemOpen
  \bibfield  {author} {\bibinfo {author} {\bibfnamefont {Y.}~\bibnamefont
  {Akamatsu}},\ }\href {\doibase 10.1016/j.ppnp.2021.103932} {\bibfield
  {journal} {\bibinfo  {journal} {Prog. Part. Nucl. Phys.}\ }\textbf {\bibinfo
  {volume} {123}},\ \bibinfo {pages} {103932} (\bibinfo {year} {2022})},\
  \Eprint {http://arxiv.org/abs/2009.10559} {arXiv:2009.10559 [nucl-th]}
  \BibitemShut {NoStop}%
\bibitem [{\citenamefont {{X. Yao and T. Mehen}}(2019)}]{yao20191}%
  \BibitemOpen
  \bibfield  {author} {\bibinfo {author} {\bibnamefont {{X. Yao and T.
  Mehen}}},\ }\href {\doibase 10.1103/PhysRevD.99.096028} {\bibfield  {journal}
  {\bibinfo  {journal} {Phys. Rev. D}\ }\textbf {\bibinfo {volume} {99}},\
  \bibinfo {pages} {096028} (\bibinfo {year} {2019})}\BibitemShut {NoStop}%
\bibitem [{\citenamefont {Yao}\ and\ \citenamefont
  {Mehen}(2021)}]{Yao:2020eqy}%
  \BibitemOpen
  \bibfield  {author} {\bibinfo {author} {\bibfnamefont {X.}~\bibnamefont
  {Yao}}\ and\ \bibinfo {author} {\bibfnamefont {T.}~\bibnamefont {Mehen}},\
  }\href {\doibase 10.1007/JHEP02(2021)062} {\bibfield  {journal} {\bibinfo
  {journal} {JHEP}\ }\textbf {\bibinfo {volume} {02}},\ \bibinfo {pages} {062}
  (\bibinfo {year} {2021})},\ \Eprint {http://arxiv.org/abs/2009.02408}
  {arXiv:2009.02408 [hep-ph]} \BibitemShut {NoStop}%
\end{thebibliography}%
\end{document}